\documentclass[preprint, 1p, 12pt]{elsarticle}

\usepackage{graphicx}
\usepackage{dcolumn}
\usepackage{bm}
\usepackage{epsfig}
\usepackage{float} 
\usepackage{amsmath}
\usepackage{amssymb}
\usepackage{subfig}
\usepackage{pifont}
\usepackage{natbib}
\usepackage{comment}
\usepackage{longtable}
\usepackage{url}

\begin{document}

\begin{frontmatter}

\title{The PHENIX Forward Silicon Vertex Detector}

\author[LANL]{C. Aidala\fnref{fn1}}
\author[UNM]{L. Anaya}
\author[LBNL]{E. Anderssen}
\author[FNAL]{A. Bambaugh}
\author[UNM]{A. Barron}
\author[LANL]{J. G. Boissevain}
\author[NMSU,UNM]{J. Bok}
\author[BNL]{S. Boose}
\author[LANL]{M. L. Brooks\corref{cor}}
\author[UNM,LANL]{S. Butsyk}
\author[LBNL]{M. Cepeda}
\author[LANL]{P. Chacon}
\author[LANL]{S. Chacon}
\author[LANL]{L. Chavez}
\author[LANL]{T. Cote}
\author[BNL]{C. D'Agostino}
\author[UNM]{A. Datta}
\author[UNM]{K. DeBlasio}
\author[FNAL]{L. DelMonte}
\author[BNL]{E. J. Desmond}
\author[LANL]{J. M. Durham}
\author[UNM]{D. Fields}
\author[Charles]{M. Finger}
\author[FNAL]{C. Gingu}
\author[FNAL]{B. Gonzales}
\author[BNL]{J. S. Haggerty}
\author[FNAL]{T. Hawke}
\author[LANL]{H. W. van Hecke}
\author[FNAL]{M. Herron}
\author[FNAL]{J. Hoff}
\author[LANL]{J. Huang}
\author[LANL]{X. Jiang}
\author[LBNL]{T. Johnson}
\author[FNAL]{M. Jonas}
\author[LANL]{J. S. Kapustinsky}
\author[UNM]{A. Key}
\author[LANL]{G. J. Kunde}
\author[UNM]{J. Kurtz}
\author[BNL]{J. LaBounty}
\author[LANL]{D. M. Lee}
\author[LANL]{K. B. Lee}
\author[LANL]{M. J. Leitch}
\author[BNL]{M. Lenz}
\author[BNL]{W. Lenz}
\author[LANL]{M. X. Liu}
\author[BNL]{D. Lynch}
\author[BNL]{E. Mannel}
\author[LANL]{P. L. McGaughey}
\author[NMSU]{A. Meles}
\author[Col]{B. Meredith}
\author[FNAL]{H. Nguyen}
\author[BNL]{E. O'Brien}
\author[BNL]{R. Pak}
\author[NMSU]{V. Papavassiliou}
\author[NMSU]{S. Pate}
\author[Sac]{H. Pereira}
\author[NMSU]{G. D. N. Perera}
\author[UNM]{M. Phillips}
\author[BNL]{R. Pisani}
\author[BNL]{S. Polizzo}
\author[Hytec]{R. J. Poncione}
\author[Prague]{J. Popule}
\author[LANL]{M. Prokop}
\author[BNL]{M. L. Purschke}
\author[LANL]{A. K. Purwar}
\author[FNAL]{N. Ronzhina}
\author[LANL]{C. L. Silva}
\author[Charles]{M. Slune\v{c}ka}
\author[Hytec]{R. Smith}
\author[LANL]{W. E. Sondheim}
\author[UNM]{K. Spendier}
\author[LANL]{M. Stoffer}
\author[NMSU]{E. Tennant}
\author[UNM]{D. Thomas}
\author[Prague]{M. Tom\'{a}\v{s}ek}
\author[Col]{A. Veicht}
\author[Prague]{V. Vrba}
\author[NMSU]{X. R. Wang}
\author[NMSU]{F. Wei}
\author[Col]{D. Winter}
\author[FNAL]{R. Yarema}
\author[LANL]{Z. You}
\author[UNM]{I. Younus\fnref{fn2}}
\author[UNM]{A. Zimmerman}
\author[FNAL]{T. Zimmerman}

\address[BNL]{Brookhaven National Laboratory, Upton, NY 11973-5000, USA}
\address[Charles]{Charles University, Ovocn´y trh 5, Praha 1, 116 36, Prague, Czech Republic}
\address[Col]{Columbia University, New York, NY 10027 and Nevis Laboratories, Irvington, NY 10533, USA}
\address[Sac]{Dapnia, CEA Saclay, F-91191, Gif-sur-Yvette, France}
\address[FNAL]{Fermi National Accelerator Laboratory, Batavia, IL 60510, USA}
\address[Hytec]{Hytec Inc., Los Alamos, NM 87545 USA}
\address[Prague]{Institute of Physics, Academy of Sciences of the Czech Republic, Na Slovance 2, 182 21 Prague 8, Czech Republic}
\address[LBNL]{Lawrence Berkeley National Laboratory, Berkeley, CA 94720, USA}
\address[LANL]{Los Alamos National Laboratory, Los Alamos, NM 87545, USA}
\address[UNM]{Department of Physics and Astronomy, University of New Mexico, Albuquerque, NM 87131, USA}
\address[NMSU]{New Mexico State University, Las Cruces, NM 88003, USA}

\cortext[cor]{Corresponding author. {E-mail address}: mbrooks@lanl.gov}
\fntext[fn1]{Present address: Physics Department, University of Michigan, Ann Arbor, MI 48109-1040, USA }
\fntext[fn2]{Present address: Physics Department, Lahore University of Management Sciences, Lahore, Pakistan }

\begin{abstract}
A new silicon detector has been developed to provide
the PHENIX experiment with precise charged particle tracking
at forward and backward rapidity.  The Forward Silicon Vertex Tracker
(FVTX) was installed in PHENIX prior to the 2012 run period of the
Relativistic Heavy Ion Collider (RHIC). The FVTX is composed of two annular endcaps, 
each with four stations of silicon mini-strip sensors, covering a rapidity range of
$1.2<|\eta|<2.2$ that closely matches the two existing PHENIX muon
arms. Each station consists of 48 individual silicon sensors, each of which
contains two columns of mini-strips with
75~$\mu$m pitch in the radial direction and lengths in the $\phi$
direction varying from 3.4 mm at the inner radius to 11.5 mm at the
outer radius.  The FVTX has approximately 0.54
million strips in each endcap. These are read out with FPHX chips,
developed in collaboration with Fermilab, which are wire bonded directly
to the mini-strips.  The maximum strip occupancy reached in central Au-Au
collisions is approximately 2.8\%.  The precision tracking provided by 
this device makes the identification of muons from secondary
vertices away from the primary event vertex possible.  The expected distance of
closest approach (DCA) resolution of 200~$\mu$m or better for particles with a transverse 
momentum of 5 GeV/$c$ will allow identification of muons from relatively long-lived particles, such as $D$ and $B$ mesons, 
through their broader DCA distributions.  

\end{abstract}


\end{frontmatter}

\section{Introduction}
\label{sec:Intro} 

A new silicon tracking detector has been developed and installed in
the PHENIX detector at the Relativistic Heavy Ion Collider (RHIC).
Since RHIC began operations in 2000, it has provided collisions of
$p$, $d$, Cu, Au, and U nuclei at center of mass energies up to
$\sqrt{s_{NN}} = 200 $ GeV for heavy ions and 510 GeV for protons.
Measurements of the matter formed at RHIC have given evidence that a
phase transition to a deconfined state of nuclear matter, the quark
gluon plasma, is achieved in ultrarelativistic collisions of large
nuclei \cite{PHENIXwhite,white2,white3,white4}.  In addition, RHIC is the world's only
collider capable of providing polarized $p+p$ collisions, which can
give information on the substructure and spin decomposition of the
proton \cite{SpinReview}.

\begin{figure}[htbp]
	\centering
	\includegraphics[scale=0.35]{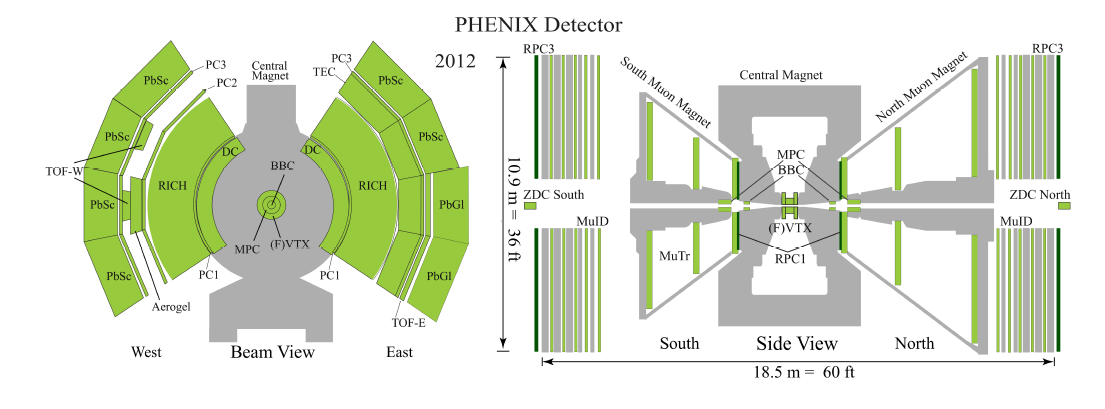}
  \caption{(color online) The 2012 PHENIX detector configuration.}
	\label{fig:Phenix_2012}
\end{figure}

The PHENIX experiment is a multipurpose detector stationed at the 8
o'clock position on the RHIC ring \cite{PHENIXNIM}.  The two central
spectrometer arms of the PHENIX detector can make measurements of
photons and electrons, as well as identified pions, kaons, and
protons, while the forward and backward arms contain specialized
detectors for tracking and identifying muons.  The existing muon arms
cover the full azimuthal angle over a pseudorapidity range of
$1.2<|\eta|<2.2$, and consist of a hadron absorber in front of three
cathode-strip chambers for charged particle tracking, followed by
alternating layers of steel absorber and Iarocci streamer tubes that
function as muon identifiers.  Finally, resistive plate chambers (RPCs) 
provide further muon identification and a fast muon trigger.
A diagram of the 2012 PHENIX detector
configuration is shown in Fig. \ref{fig:Phenix_2012}.

Production of heavy charm and bottom quarks at forward rapidity is of
particular interest to the RHIC heavy ion and spin physics programs.
Open heavy flavor hadrons at forward angles are measured at PHENIX
through their semi-leptonic decays to muons; however, such
measurements are limited by systematic uncertainties due to the large
backgrounds from $\pi^{\pm}$ and $K^{\pm}$ decay muons, and hadrons
which penetrate the absorbers \cite{PPG117, PPG153}.  Also, without precise
knowledge of the event and decay vertices, the contributions from $D$
and $B$ meson decays cannot be separated.  Quarkonium measurements in
the dimuon channel suffer from large combinatorial backgrounds in A+A
collisions, which limit the precision of the extracted yields.  Since
the tracking stations are located behind the hadron absorber, muons undergo
multiple scattering in the absorber material before any measurement
occurs.  This scattering affects the measured opening angle of the 
muon pairs, which in turn degrades the mass resolution of dimuon resonances.
				   
The Forward Silicon Vertex Detector (FVTX) enhances the
capabilities of the existing PHENIX muon arms by adding additional
tracking in front of the hadron absorbers. The four tracking stations 
of the FVTX add precision vertexing capability
to the existing muon spectrometers, with full azimuthal coverage and, when 
combined with the central silicon vertex detector \cite{VTX}, a rapidity coverage 
of $1.2<|\eta|<2.2$.  By identifying tracks which
originate away from the primary interaction vertex, the FVTX
enables rejection of muons which result from decays of relatively long lived
particles like $\pi^{\pm}$ , $K^{\pm}$, and $K^{0}_{L}$ 
when searching for heavy quark and $W^{\pm}$ decay muons.  In
addition, muons from decays of hadrons containing open heavy flavor are identifiable
through their broad DCA distributions, and
prompt muon pairs from the Drell-Yan process can be tagged as
originating from the primary collision vertex.  With precision
tracking in front of the absorber, the opening angle of muon pairs
can be measured before any additional scattering occurs, giving a
more precise dimuon mass resolution and enabling the separation of the
$\psi(2s)$ peak from the larger $J/\psi$ peak in the dimuon mass
spectrum.  In A+A collisions, the FVTX can also be used to determine the
reaction plane.

This paper presents a comprehensive report on the design,
construction, and first operation of the FVTX.  A general over view of the detector 
is given in Section \ref{sec:overview}. Section \ref{sec:electronics} describes the sensor 
and electronics associated with processing and reading out the data.  Section \ref{sec:mechanical} details
the mechanical design of the detector and related infrastructure, and explains how various detector 
components were assembled. The effects of radiation on the FVTX silicon sensors are discussed in 
Section \ref{sec:radiation}, and Section \ref{sec:performance} shows the initial performance 
of the detector during RHIC's 2012 and 2013 run periods.  Finally, a summary 
of the paper is given is Section \ref{sec:conclusions}.
 
\section{Detector Overview}
\label{sec:overview}

\begin{figure}[htbp]
  \centering
  \includegraphics[width=0.6\linewidth]{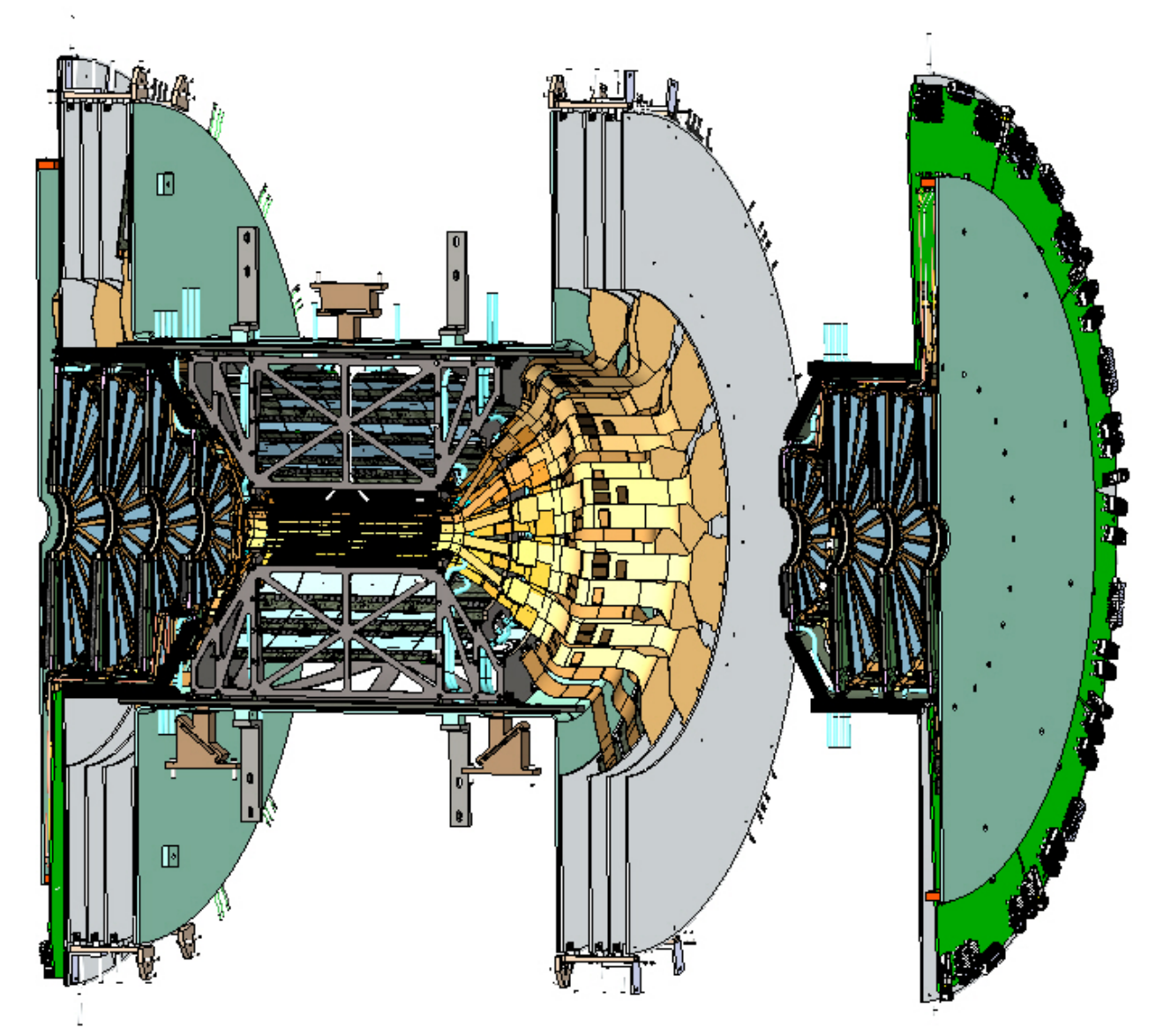}
  \caption{(color online) A drawing of the combined VTX/FVTX assembly.  One FVTX
    quadrant is displaced for clarity.}
  \label{fig:FVTX_exploded}
\end{figure}

The FVTX detector system is composed of two identical endcap sections,
located on either end of a 4-layer barrel silicon vertex
detector (VTX) \cite{VTX}, and in front of the north and south muon
spectrometer arms.  Each endcap has 4 layers of active
silicon sensors arranged in a disk around the beryllium beam pipe. 
The basic unit of construction is a \emph{wedge}, (section
\ref{subsec:wedges}) each of which carries a mini-strip silicon
sensor (section \ref{sec:sensor}), read-out chips (section \ref{sec:chip}), and a high-density interconnect (HDI, section \ref{sec:hdi}). 
Wedges are mounted on \emph{half-disks} (section
\ref{subsec:disks}), and fitted with extension cables (section
\ref{sec:extension}). For simplicity, the half-disks are referred to as
\emph{disks}.  Disks are mounted into \emph{cages} (section \ref{subsec:cage}), and
the extension cables are connected to \emph{ROC boards} (section
\ref{sec:roc}), the read-out cards that are the first stage in the
data path.  Finally the cage+ROC assembly is installed in the
carbon-composite frame which also contains the VTX components.

Fig. \ref{fig:FVTX_exploded} shows a model of two quadrants of the detector, with one
FVTX quadrant displaced in $z$ for clarity. The
wedges, mounted on the disks, are shown installed into their cages.  The VTX and its
associated electronics are shown in the middle, mounted in
the support frame. As can be seen in the figure, each cage has one
small and three large disks.  The smaller disks are simply truncated 
versions of the larger disks.  A summary of the FVTX design parameters is given in Table \ref{tab:mech_summary}.

\begin{table}
  \caption{Summary of design parameters}
  \label{tab:mech_summary}
  \begin{tabular}[htpp]{|l|c|}
                                                                                 \hline
Silicon sensor thickness  ($\mu$m)   & 320                                                 \\ \hline
Strip pitch  ($\mu$m)   & 75                                                 \\ \hline
Nominal operating sensor bias (V) & +70                                   \\ \hline
Strips per column for small, large wedges & 640, 1664                    \\ \hline
Inner radius of silicon (mm) &  44.0                              \\ \hline
Strip columns per half-disk (2 per wedge) & 48               \\ \hline
Mean z-position of stations (mm)  & $\pm$201.1, $\pm$261.4, $\pm$321.7, $\pm$382.0 \\ \hline
Silicon mean z offsets from station center (mm) &  $\pm$5.845, $\pm$9.845         \\ \hline

  \end{tabular}
\end{table}

\section{Electronics}
\label{sec:electronics}

This section describes the electrical components and support
systems used to read out and power the FVTX. The silicon mini-strip sensors
and the FPHX read-out chips are described in sections \ref{sec:sensor} and \ref{sec:chip}, respectively. 
The HDI that provides power, bias voltage, and slow control signals to the sensor is discussed in section 
\ref{sec:hdi}, and the extension cables which handle signals between the wedges and the read-out cards are described in 
section \ref{sec:extension}.  The read-out cards and front end modules which process signals from the detector are discussed
in section \ref{sec:roc}, followed by a description of the high- and low-voltage delivery systems in section \ref{sec:HV_LV}.

\subsection{Sensors}
\label{sec:sensor}

\begin{figure}[htbp]
  \centering
  \includegraphics[width=0.8\linewidth]{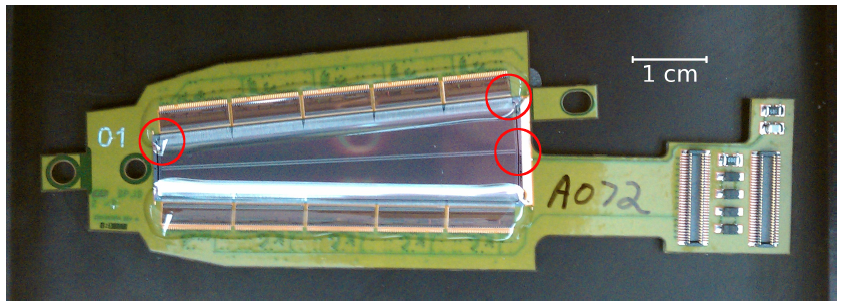}
  \caption{(color online) A completed FVTX small wedge, with sensor facing up.  Note the center line dividing the
two halves of the sensor and rows of FPHX chips along the sensor edges.}
  \label{fig:wedge_photo}
\end{figure}

The silicon mini-strip sensors were designed at Los Alamos and
fabricated by Hamamatsu Photonics KK. The wedge-shaped geometry comprises two
individual columns of strips that are mirror images about the center
line on the same sensor.  The wire bond connections between the strips
and read-out chips are located along the outer edges of the sensor
(see Fig. \ref{fig:wedge_photo}).  The centerline gap between
columns is 100~$\mu$m and is completely active.

\begin{figure}[htbp]
  \centering
  \includegraphics[width=1.0\linewidth]{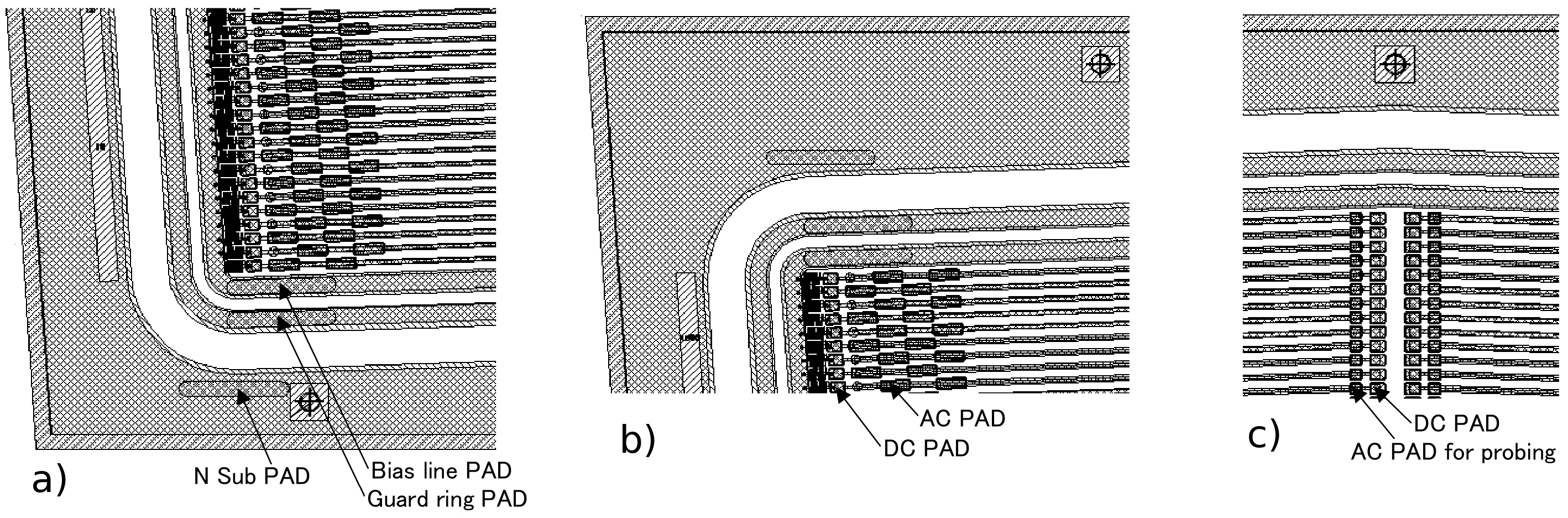}
  \caption{Details of the sensor layout. a) Narrow end corner, b) wide end corner and
c) wide end center. These areas correspond to the circled regions in Figure \ref{fig:wedge_photo}.}
  \label{fig:sensor_detail}
\end{figure}

The strip length increases with radius on the sensor, and goes from 3.4~mm at the inner
radius to 11.5~mm at the outer radius, with a pitch of 75~$\mu$m in the radial direction. Each sensor
covers 7.5$^\circ$ in $\phi$, and since the strips are perpendicular to the radius, they make
an angle of 86.25$^\circ$ with respect to the centerline, as can be seen in Fig. 
\ref{fig:sensor_detail}.

The sensors were fabricated with p-implants on a 320~$\mu$m thick n-type
substrate.  The strips are AC-coupled and biased through individual
1.5~M$\Omega$ polysilicon resistors to a typical operating voltage of +70~V.  
The metallization on the strips is wider than the implant to provide field 
plate protection against micro-discharges, a concern that becomes greater with
radiation-induced increases in the leakage current.  The strips are
also protected by two p-implant guard rings and an n$^+$ surround
between the guard rings and sensor edge.  There are two sets of bond
pads for each strip, one of which is dedicated to probe tests.  Each
strip also has a spy pad, which is an opening through the capacitor
oxide layer, that allows the DC characteristics of the strip to be probed.
Fig. \ref{fig:sensor_detail} shows details of the sensor layout, including 
guard rings, bond pad locations, and mechanical fiducial marks.

\subsection{FPHX Chip}
\label{sec:chip}

\begin{figure}[htbp]
  \centering
  \includegraphics[trim =0mm 2mm 0mm 0mm, clip,width=1.0\linewidth]{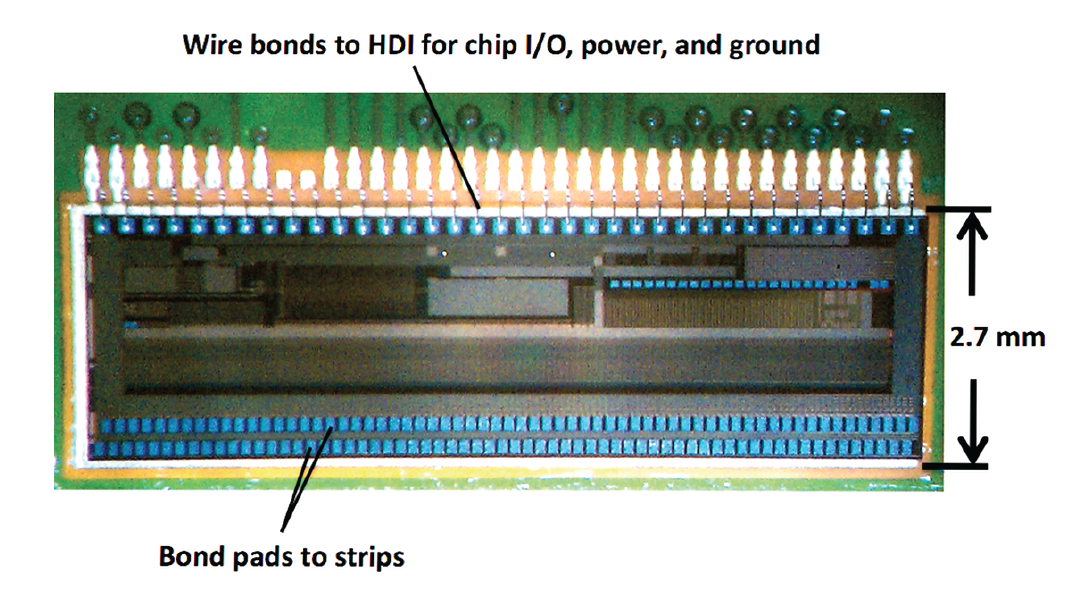}
  \caption{A single FPHX chip mounted onto the HDI. Along the top, wire bonds to the HDI have been completed. Along the bottom, in two rows, 
are 128 bond pads for wire bonds to the silicon strips.}
  \label{fig:chip_zoom}
\end{figure}

A custom 128-channel front-end ASIC, the FPHX, was designed by Fermilab
for the FVTX detector \cite{FPHX, FPHX_JonK}. The chip was optimized for fast trigger
capability, a trigger-less data push architecture, and low power
consumption. The chip was fabricated by the Taiwan Semiconductor
Manufacturing Company (TSMC) with 0.25 $\mu$m CMOS technology. The
analog section consists of an integrator/shaper stage followed by a
three-bit ADC. A single FPHX chip mounted onto the HDI is shown in Fig. \ref{fig:chip_zoom}.  
In this example, the wire bonding to the control lines on the HDI is complete, 
but no bonding between the sensor and the chip has been performed.

Many of the setup parameters of the front end are
programmable via an LVDS serial slow control line. Adjustable
parameters include gain, threshold, rise and fall time, input
transistor bias current, channel mask, plus several additional fine
tuning parameters. Nominal setup parameter values include integrator/shaper
gain of 100~mV/fC, 800 mV dynamic range, 60 ns risetime at the
shaper output, and a 2000-electron threshold at the first
comparator. All analog functions were exercised on a test stand and
the measured values were very close to pre-submission simulations. 

The FPHX chip was designed to process up to four hits within four RHIC
beam crossings (or $\sim$4$\times$106~ns = 424~ns). Each hit contains a 7-bit
time stamp, 7-bit channel identifier, and a three-bit ADC value. By only accepting 
hits above a certain (programmable) ADC threshold, the signal-to-noise ratio can be dynamically 
optimized for different operating conditions.
In addition, the ADC information from strips in an FVTX hit cluster is used to determine 
the center of the track via a weighted average of the charge in each strip.  An ADC with 
higher resolution would not significantly improve the detector's tracking resolution, 
since multiple scattering in detector material is the dominant contribution to track 
smearing at the $\sim$20~$\mu$m level.

The data words are output over two LVDS serial lines at up to 200~MHz
clock rate. The total power consumption of the FPHX is $\sim$390
$\mu$W per channel. The noise, when the chip was wire bonded to a
sensor with strips $\sim$2--11~mm in length ($\sim$1-2.5~pF) was
simulated and measured to be below the design specification of $500$ electrons.

\subsection{High-Density Interconnects}
\label{sec:hdi}

\begin{figure}[h]
  \centering
  \includegraphics[trim=0cm 4cm 0cm 4cm, clip=true,width=0.8\linewidth]{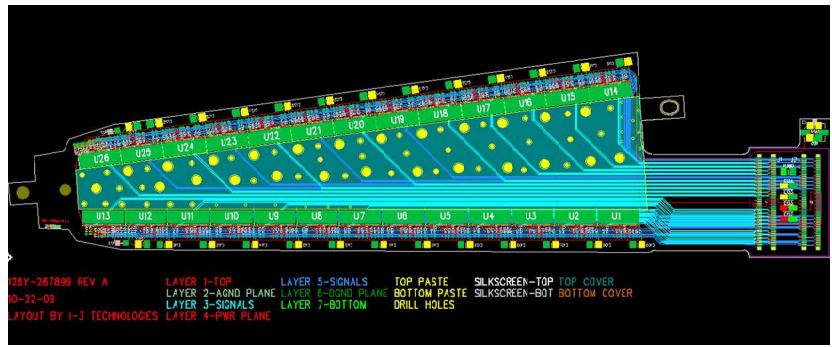}
  \caption{(color online) Schematic of the HDI stack-up.}
  \label{fig:HDIstackup}
\end{figure}

The silicon sensor and FPHX read-out chips are assembled on an 
HDI which provides the slow control,
power, and bias input lines as well as slow control and data output
lines.  The HDI stack-up is shown in Fig.~\ref{fig:HDIstackup} and
consists of seven layers of single- sided (20~$\mu$m) and double-sided
(50~$\mu$m) copper coated polyamide bonded together with a 25~$\mu$m
sheet adhesive for a total thickness of approximately 350~$\mu$m. Indicated on the HDI
stack-up are two signal layers, one ground layer, and one power layer.  All
control lines (which are not active during data taking) are routed
under the sensor, and all output lines are routed towards the edge of
the wedge, thus minimizing the coupling between the output lines and
the sensor.  The number of lines (8 pairs for the control
lines and 2 signal pairs per chip for the output lines) requires that they
 have a 40~$\mu$m width with a 100~$\mu$m spacing. Both
simulated and physical tests were carried out to ensure that the input
clock (200 MHz) had sufficient integrity at the furthest point from
the driver.

Since the layout of the wedge, chip, and HDI can impact the system
noise, the electrical layout of the wedge assembly was designed to
minimize any additional noise.  Incorporated onto the HDI were two
noise-canceling loops, one for the input side and one for the output
side of the chip, employing bypass capacitors connected to the bias
ground and digital ground, respectively.  Termination resistors for the
calibration lines and bias resistors and capacitors are also located
on the board.  All elements are surface mounted and assembled by the
manufacturer (Dyconex and Micro Systems Engineering).

\subsection{Extension cables}
\label{sec:extension}

The extension cables were designed to bring all signals from the HDI
to the ROC board and power from the ROC board to the wedge, and have a
similar stack-up design to the HDIs.  The mechanical layout of the
extension cables was unique for each side and each disk of the FVTX to
allow the ROC end of the extension cables to precisely line up to the
corresponding connector on the ROC.  Both the HDI and the extension
cables needed to be permanently bent in multiple directions in order
to precisely fit them within the constraints of the detector envelope.
Bending was accomplished using fixtures which bent the flex circuits and 
heated them at $100^\circ$C for $\sim$5 minutes, in order to introduce 
a permanent deformation.

\subsection{ROCs/FEMs}
\label{sec:roc}

The design of the read-out electronics for the FVTX detectors is based
on three major constraints, imposed by the detector:
\begin{itemize}
\item Large instantaneous bandwidth (3.38 Tb/s)
\item Radiation hardness of read-out components near the interaction
  point
\item Large number of I/O lines (21,000 LVDS pairs)
\end{itemize}

\begin{figure}[htbp]
  \centering
  \includegraphics[width=1.0\linewidth]{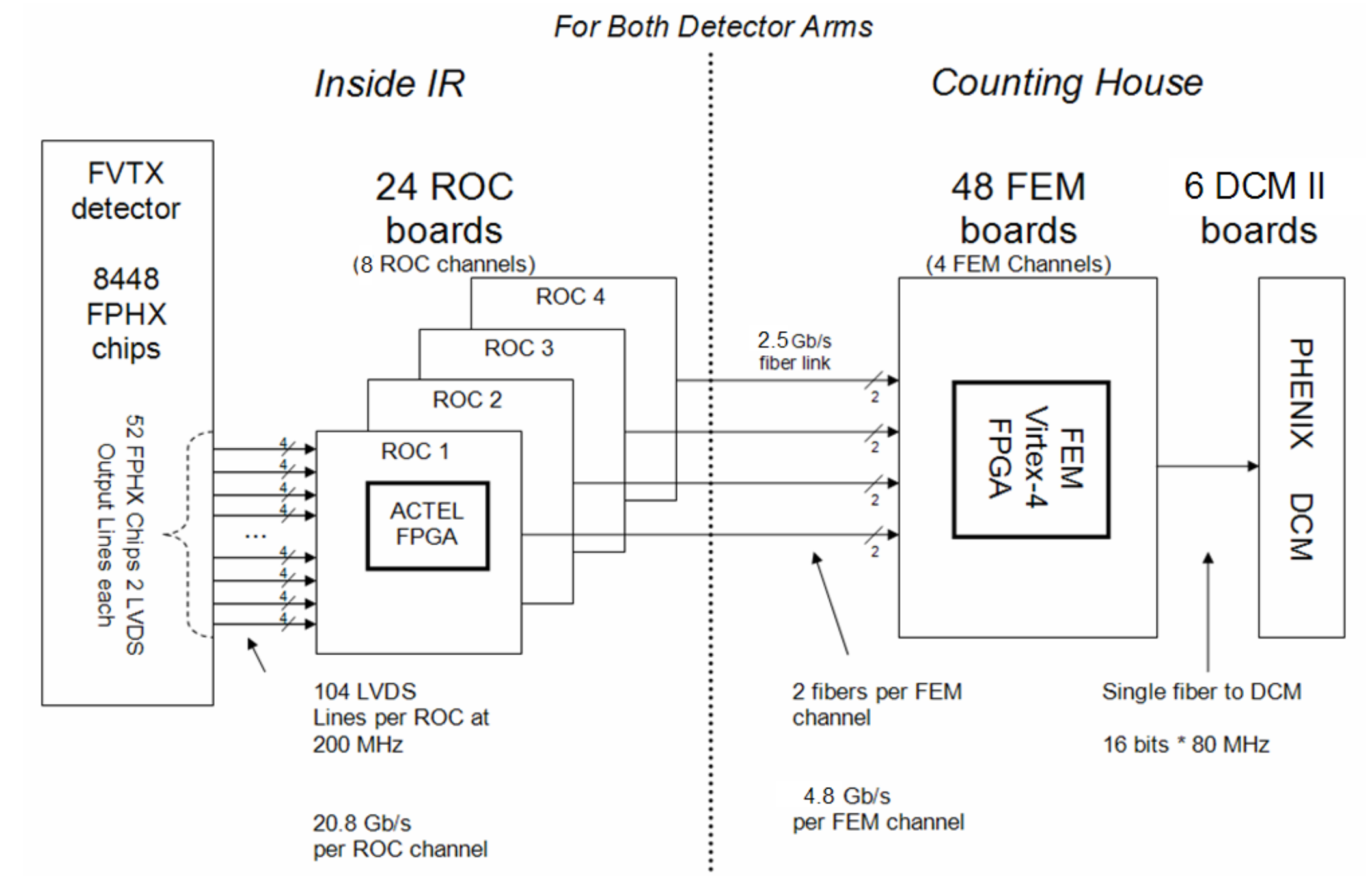}
  \caption{Read-out electronics block diagram.}
  \label{fig:fvtxreadout}
\end{figure}

As a result, the read-out electronics are logically divided into two
independent blocks, illustrated in Fig.~\ref{fig:fvtxreadout}. The
components are:
\begin{itemize}
\item Read Out Card (ROC) - module which is located close to the
  detector.
\item Front End Module (FEM) - module which is located in the
Counting House ($\sim$50~m from the Interaction Region) in a standard
VME crate.
\item FEM Interface Board - module located in each of the FEM VME
crates.
\end{itemize}
The output of the FEM connects to the standard PHENIX DAQ board, a Data Collection Module (DCM),
and from this point on the data stream becomes a part of the standard PHENIX DAQ.

\subsubsection{ Read Out Card (ROC)}
The ROC boards are mounted on an aluminum cooling plate and connected to the HDIs through
the extension cables. Since the board have components on both the top and bottom surfaces, a 1/8 in
thick layer of soft non-conductive Gap-Pad material is placed 
between the ROC boards and the cooling plate in order to facilitate heat transfer.

 The ROC boards are designed to:
\begin{itemize}
\item Receive data via LVDS pairs from the silicon read-out chips
\item Combine and synchronize the data streams from multiple FPHX chips
\item Send the data to the front end module (FEM) in the counting house via optical fibers
\item Receive and distribute slow control data to/from the FPHX chips
and other ROC components
\item Hold an on-board calibration system for the FPHX chips.
\item Hold an on-board JTAG FPGA which allows for remote programming of the slow control and data FPGAs.
\end{itemize}

\begin{figure}[htbp]
  \centering
  \includegraphics[width=0.6\linewidth]{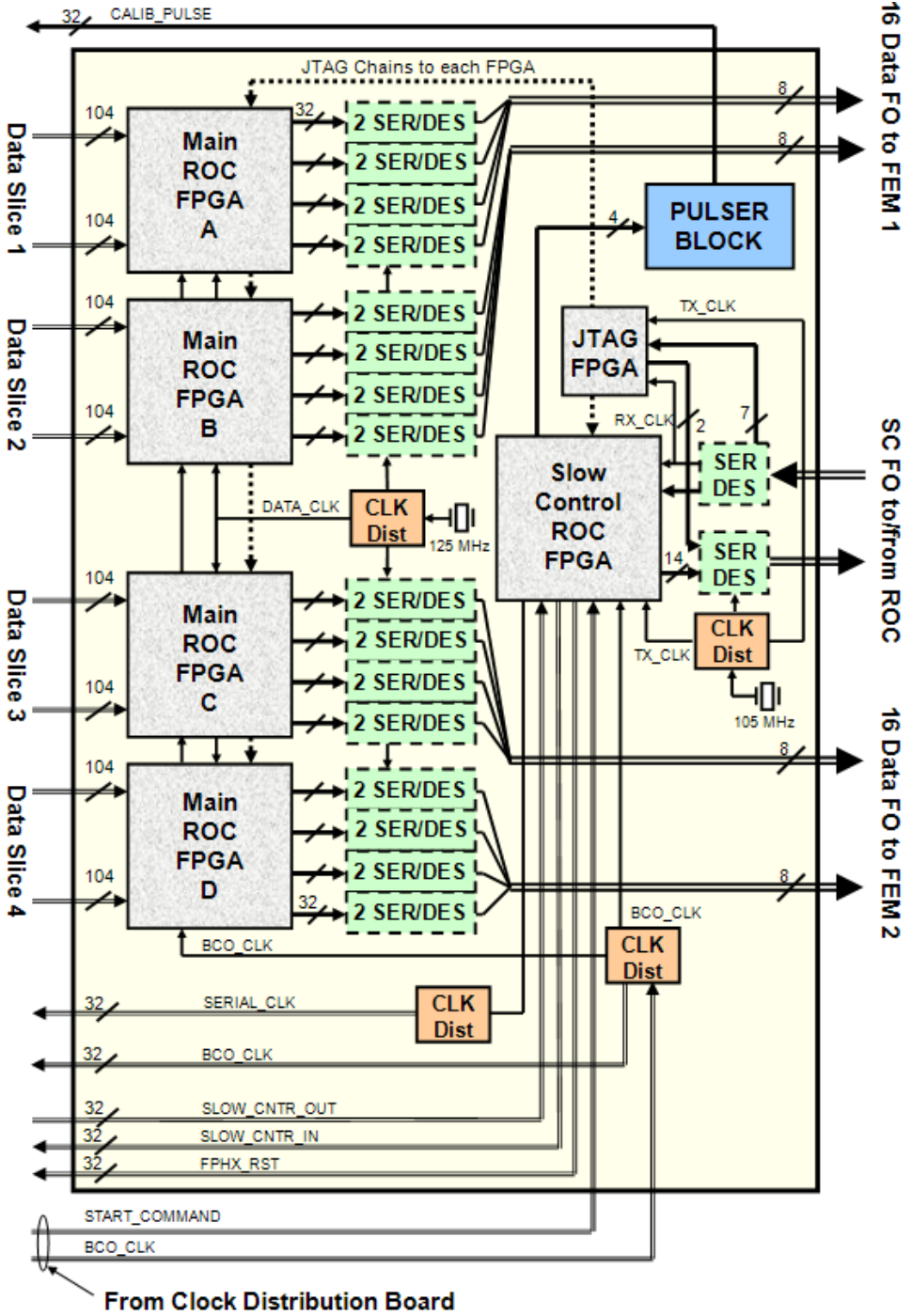}
  \caption{(color online) Block diagram of the ROC board.}
  \label{fig:rocdiagram}
\end{figure}

The ROC board design utilizes radiation-hard FLASH-based ACTEL
ProASIC3E FPGAs in order to limit susceptibility to single event
upsets (SEUs). A functional ROC board diagram is shown in
Fig.~\ref{fig:rocdiagram}. Each board contains 4 large-scale ACTEL
A3PE3000-FG896 FPGAs to process the data from the read-out chips, 33
16-bit Serializer/Deserializer chips (TLK2711) and four 12-channel
optical fiber transmitters (HFBR-772BEZ) to send the data to 2 FEM
boards. Each ROC FPGA holds two completely independent ROC channels,
for a total of 8 ROC channels per board, which send out 32-bit data at
the output clock frequency of 125 MHz. The outgoing data are logically
split into two hi/lo 16-bit data portions which are each sent to
separate Serializer/Deserializer chips and a single fiber channel. A
small-scale ACTEL FPGA is dedicated to distributing slow control
signals to appropriate chips and reading data back, to be sent up the
slow control data stream to the FEM. slow control data are sent by the
FEM to the ROC and by the ROC to the FEM over a single optical fiber
interface.  Another small-scale ACTEL FPGA provides an interface between 
the slow control fiber and on-board FPGAs, to allow remote programming 
of the slow control and data FPGAs.

The Beam Clock (9.4~MHz) arrives at the ROC board as an LVDS signal
and is distributed to all the FPGAs on the board as well as to all
the FPHX chips. A Serial Clock of 20$\times$ the Beam Clock frequency is
generated by a PLL on the slow control FPGA. The output data from the
FPHX chips are phase latched to a similarly generated 20$\times$ clock
inside the Main FPGAs, which avoids distribution of the fast
clock between FPGAs and simplifies the design.  

\begin{figure}[htbp]
  \centering
  \includegraphics[trim=0cm 3cm 0cm 0cm,clip=true, width=0.55\linewidth]{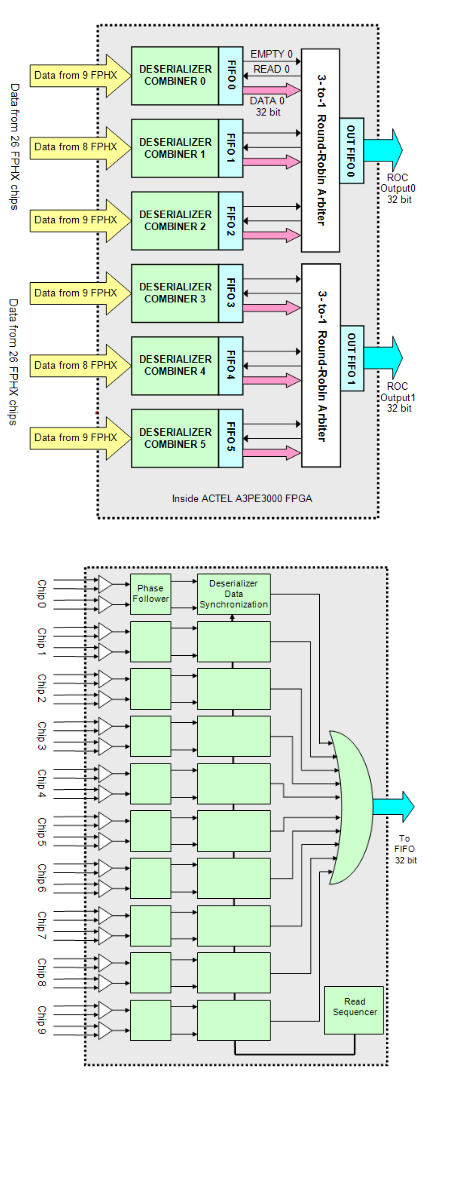}
  \caption{(color online) Block diagram of a single ROC channel and a single 10-chip
    channel deserializer/combiner.}
  \label{fig:rocchannel}
\end{figure}

A schematic of a single ROC Channel is shown in Fig.~\ref{fig:rocchannel}. 
The main task is to combine data from up to 10 FPHX chips into a single data stream
without any delay: 20-bit data deserialization gives time for
this. Three of those streams are combined by a 3-to-1
Round-Robin Arbiter and buffered into a 256 deep output FIFO. We
utilize triple redundancy on every component that allows for it, and
actively use design blocks for predictable layout and timing. The
design is latch free by construction, with constant synchronization of
the input serial data streams.

The ROC board includes a calibration system that can deliver a
precisely controlled voltage pulse to each FPHX chip on an HDI. The
signal injection timing is synchronized with the Beam Clock. A 10 bit
dual DAC is used together with a precision reference and analog
switches to provide a large dynamic range and low noise. The amplitude
is adjustable via the slow controls, while a fast rise and slow fall
time are fixed by RC circuits. The calibration system is used
routinely to check for dead FPHX channels and determine the electronics noise levels. 
When disabled, the system contributes negligible noise to the FPHX chips.

\subsubsection{ROC Radiation Environment}
The primary consideration when evaluating FPGA technology available 
for use on the ROC was the effect of the radiation
environment on the performance of the system.  Additional
considerations included I/O configurations, serial communication
capabilities, and the ability to reconfigure the device within the system.  The
choice is primarily the selection of configuration memory
technology, as logic implementation and density have minimal impacts in this 
application.  The different configuration technologies and their
suppliers that were considered are as follows:

\begin{itemize}
\item SRAM Altera, Xilinx
\item FLASH Actel ProASIC3
\item Anti-fuse	Actel Axcelerator
\end{itemize}

The primary elements of the FPGA that are affected by the radiation
are the SRAM memory elements, clocks, and sequential logic.  Since the 
configuration of Altera and Xilinx FPGAs are contained in SRAM, upsets
in this memory will affect functionality of the device. Both Xilinx and Altera
offer configuration ``scrubbing'' solutions that check the
configuration, but they require a reload of the configuration if an
error is detected, which takes time.  The Actel FPGAs do not have SRAM 
configuration memory so they are immune to this form of upset, which
is a distinct advantage.  The Actel ProASIC3 was chosen. In the 2012 and 2013
runs at RHIC, no problems due to radiation effects on the ROC FPGAs were observed.

\subsubsection{Front End Module (FEM)}

The FEM boards are located in 6U VME crates in the counting house
(in a shielded location $\sim$50 meters from the detector) where radiation levels are
negligibly small and SRAM-based FPGAs can be used.  The FEM boards are
functionally designed to:
\begin{itemize}
\item Receive data from the ROC boards over fiber links.
\item Sort the incoming data according to the Beam Clock Counter.
\item Buffer the data from the last 64 beam clocks.
\item Upon Level-1 trigger decision, ship the data from the Beam Clock of
interest to the output buffer, which ships data to the PHENIX Data
Collection Modules (DCM).
\item Distribute and receive slow control data to/from the ROC
cards. The online slow control interface is made through the FEM
Interface Board and the interface to the ROC cards is made through an
optical fiber.
\end{itemize}

\begin{figure}[htbp]
  \centering
  \includegraphics[width=0.6\linewidth]{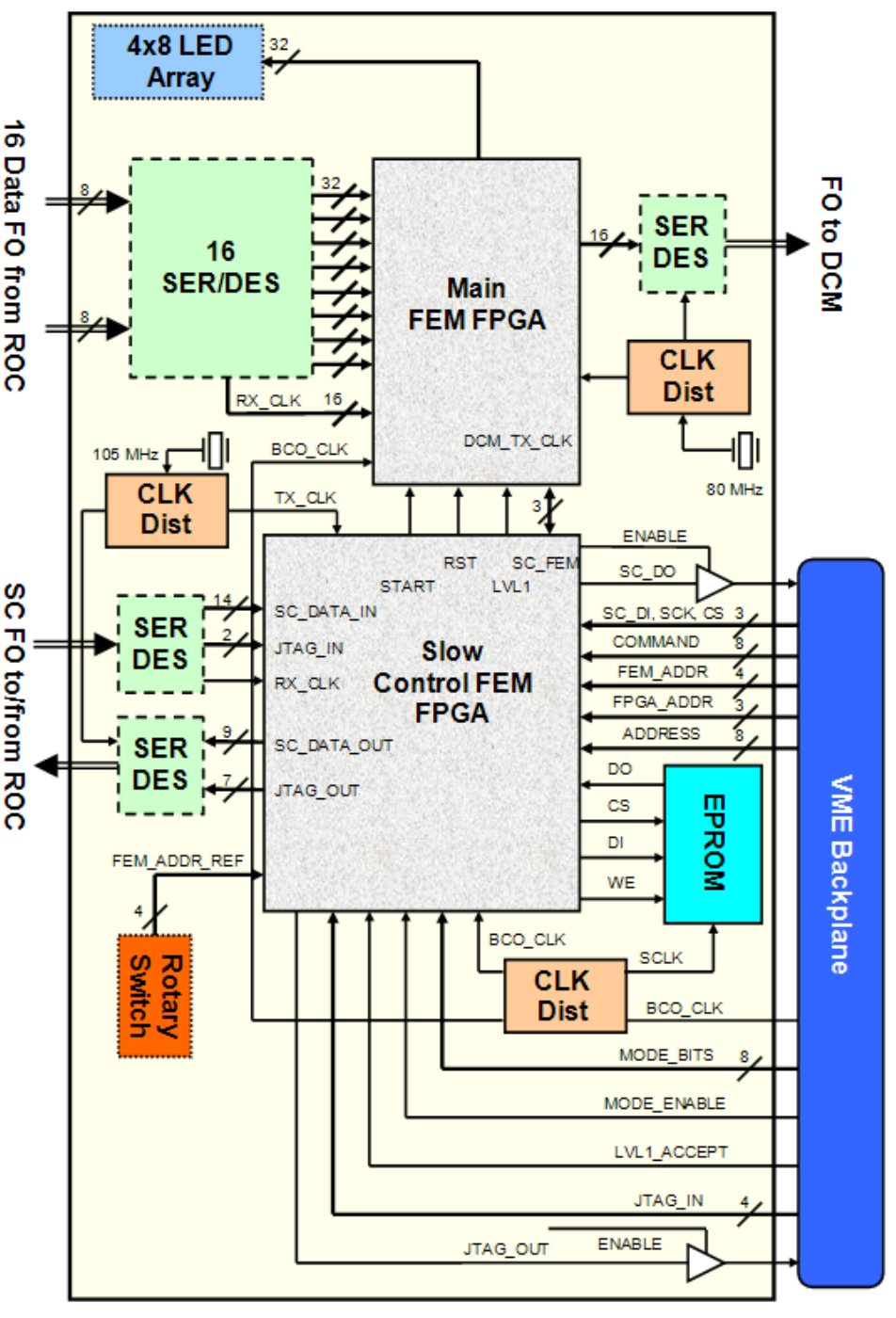}
  \caption{(color online) Block diagram of a FEM board.}
  \label{fig:femdiagram}
\end{figure}

The FEM board architecture can be seen in Fig.
\ref{fig:femdiagram}. Xilinx Virtex-4 FPGAs are used as the main FEM FPGA.  
The largest device of the memory-intense SX series
(XC4VSX55) provides enough fabric and memory to implement 4 FEM
channel cores and a channel combiner on a single FPGA. This
significantly reduced the cost of the FEM board design.

\begin{figure}[htbp]
  \centering
  \includegraphics[width=0.9\linewidth]{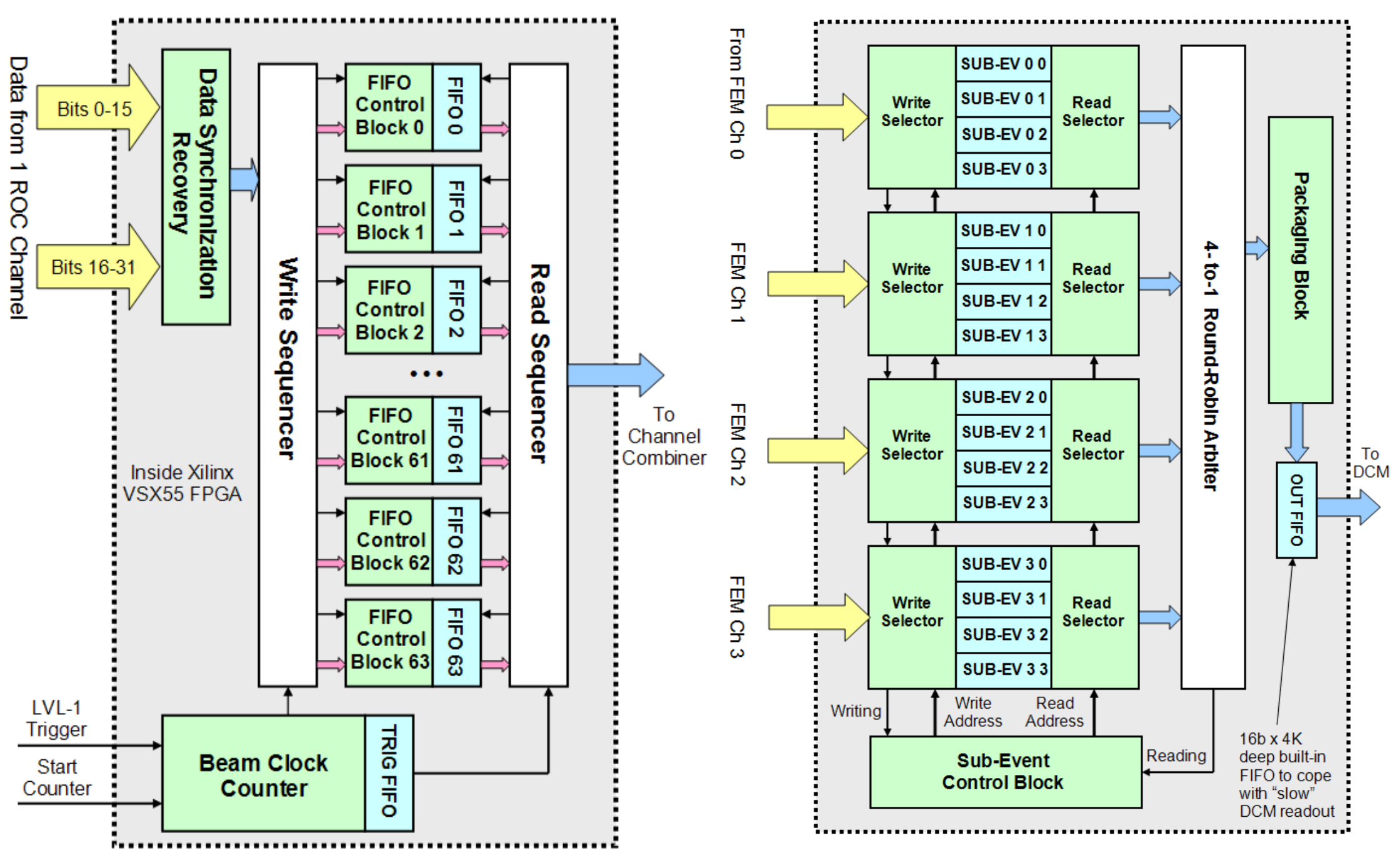}
  \caption{(color online) Block diagram of a single FEM channel and a channel
    combiner.}
  \label{fig:femchannel}
\end{figure}

A block diagram of a single FEM channel and FEM channel combiner is shown
in Fig.~\ref{fig:femchannel}. One FEM board receives 16 optical
fibers (from half of a ROC board). The incoming 16-bit data nibbles
from the ROC are aligned and combined into 32-bit data words;
alignment bits are used in transmission.  Data are buffered for 64
beam clocks in each of the 4 FEM channels in an array of 64 512 word
deep FIFOs. Each FIFO stores the data for a particular beam bucket (0
through 63). Since data from the FPHX chips carry the
7-bit beam clock counter information, the sorting is trivial.

\begin{figure}[htbp]
  \centering
  \includegraphics[width=0.6\linewidth]{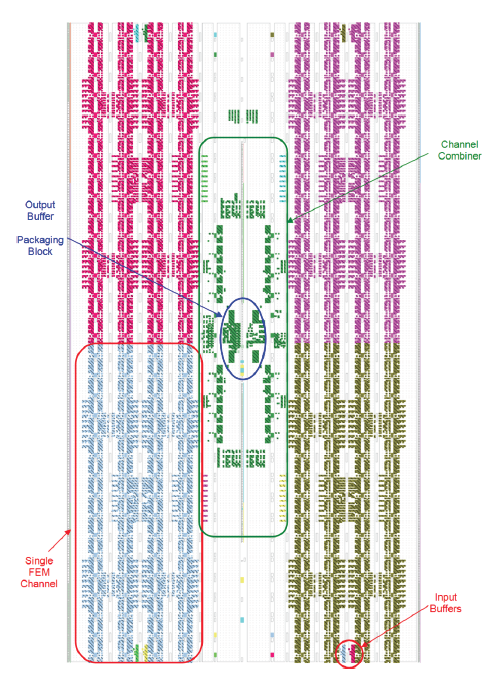}
  \caption{(color online) Implementation of the FEM FPGA design.}
  \label{fig:femfloor}
\end{figure}

Upon receiving a Level-1 trigger, the FEM channels start sending the
data from a particular beam bucket of interest to the output sub-event
buffer. The internal clock speed for reading and writing is 300~MHz in
order to minimize latencies in the design. The resulting sub-events
are combined and packetized in the ``Channel Combiner'' block into
16-bit wide packets.  Packets carry fixed data (Event Counter, BCO
Counter and Longitudinal Parity) in header and trailer words.  The
design is implemented using Xilinx RPM blocks and directed routing
concepts.  The current version of the design can run at up to 350~MHz
clock on a -10 speed grade FPGA. The Floorplanner diagram of the
current implementation is shown in Fig. \ref{fig:femfloor}.  A
separate small-scale FPGA (Spartan3 XC3S200) processes incoming timing
signals and slow control commands. Timing and fast control signals in
PHENIX are distributed through a specially designed Granular Timing
Module (GTM). The GTM signals include the Beam Clock, a Level-1
Trigger and Control Mode Bits. These signals are copied to the main
FEM FPGA and the beam clock is selected out by the FEM Interface Board
and sent to the ROC board.

\subsubsection{FEM Interface Board}

\begin{figure}[htbp]
  \centering
  \includegraphics[width=1.0\linewidth]{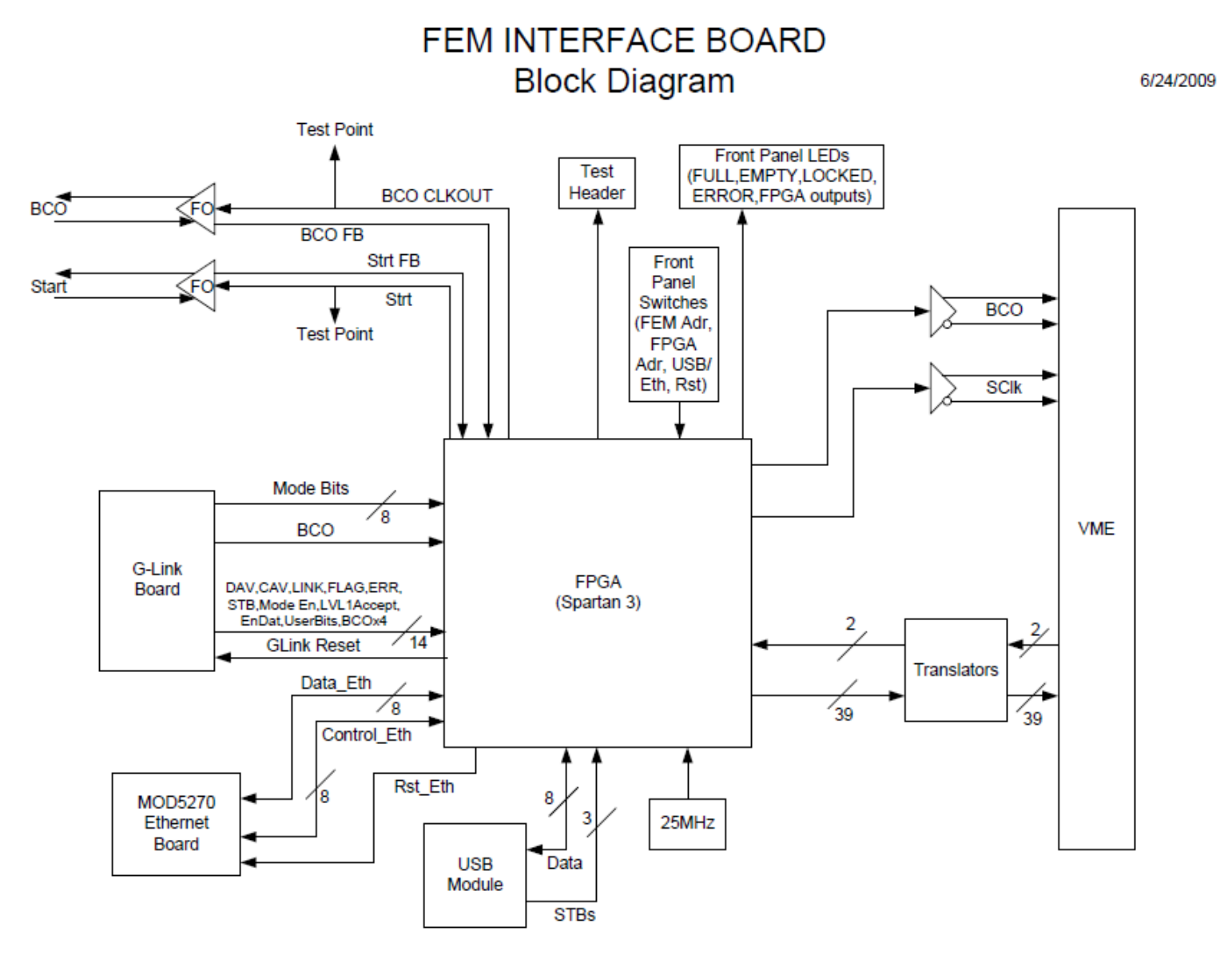}
  \caption{Block diagram of the FEM Interface Board.}
  \label{fig:femib}
\end{figure}

Each VME Crate with 12 FEM boards is controlled by a single 6U FEM
Interface Board which acts as a simple crate controller and is
designed to:
\begin{itemize}
\item Receive GTM (beam clock and trigger) signals.
\item Distribute the information from the GTM to the FEMs via the VME
backplane.
\item Distribute the Beam Clock to a set of ROCs via a front-panel
fiber optic interface.
\item Distribute a START signal to a set of ROCs via a front-panel, to
allow the FEM and ROC read-outs to be synchronized.
\item Interface to the PHENIX Online slow controls system via Ethernet
or USB, and to the FEM boards via the VME backplane.
\end{itemize}
The FEM interface board utilizes a commercial MOD5270 Motorola
Coldfire Ethernet Daughter Board and FT2232 USB Module to provide a
bi-directional slow control data stream, and also holds a PHENIX standard G-Link
Receiver board. The full block diagram of a FEM Interface Board is
shown in Fig. \ref{fig:femib}.

\subsection{Detector Bias, Low Voltage, and Grounding}
\label{sec:HV_LV}

Each detector wedge requires independent 2.5 V digital and analog
power sources as well as a bias supply. Thus, 768 DC supply channels
and 384 bias channels are needed for the full FVTX detector. In addition, 
each ROC uses a number of moderately high power DC sources to operate the boards. 
Bulk DC power is provided by Vicor MegaPACs. 
These high-power switching supplies are connected to modules in a low voltage 
DC power crate system designed at BNL. Each module
provides ten channels of up to 10A DC output, with independently
monitored voltage and current. Power for the ROC cards is wired
directly to these outputs. Low voltage distribution crates were
designed to fan out these high current channels to all of the
wedges. The power to each wedge is individually enabled via Ethernet
slow controls. The distribution cards have individual fuses per wedge
and filter each DC source and return line using pi networks.

A Wiener Mpod system serves as the bulk supply for detector bias voltage.  It
provides 48 low current channels with adjustable voltage and current
readback through Ethernet.  Another distribution crate was designed
to fan out these channels to each wedge, controlled by Ethernet. Each
output is protected against over voltage using zener diodes.  The
source and return lines are filtered as described above.  The typical 
bias voltage applied to each sensor during operation is +70~V.

\subsubsection{Grounding System}

A star topology was chosen for the overall grounding plan. At the
center of the star is each wedge/HDI. The wedge bias, FPHX analog and
digital grounds are tied together only at their respective HDIs. The
digital grounds are returned to the ROC cards through the wedge
extension cables, which are all tied together and earthed to the
PHENIX central magnet. Electrically filtered distribution boards,
described in the previous section, prevent ground loops and noise from
being introduced by the remote power and bias supplies.

\section{Mechanical Design}

\label{sec:mechanical} 

Here we describe the mechanical design of the FVTX and the procedures used to assemble the various 
components into a complete detector.  The assembly of the silicon sensors, FPHX chips, and HDI into 
a \emph{wedge} is explained in section \ref{subsec:wedges}. The \emph{disks} which hold the wedges and the 
\emph{cages} which house the disks are described in sections \ref{subsec:disks} and \ref{subsec:cage}, respectively. 
The cooling system used to remove heat generated by the detector electronics is described in section \ref{subsec:cooling}, 
and the environmental enclosure that is necessary to avoid condensation on the cooled electronics is discussed in section \ref{subsec:enclosure}.

\subsection{Wedges}
\label{subsec:wedges}

\begin{figure}[htbp]
  \centering
  \includegraphics[width=0.8\linewidth]{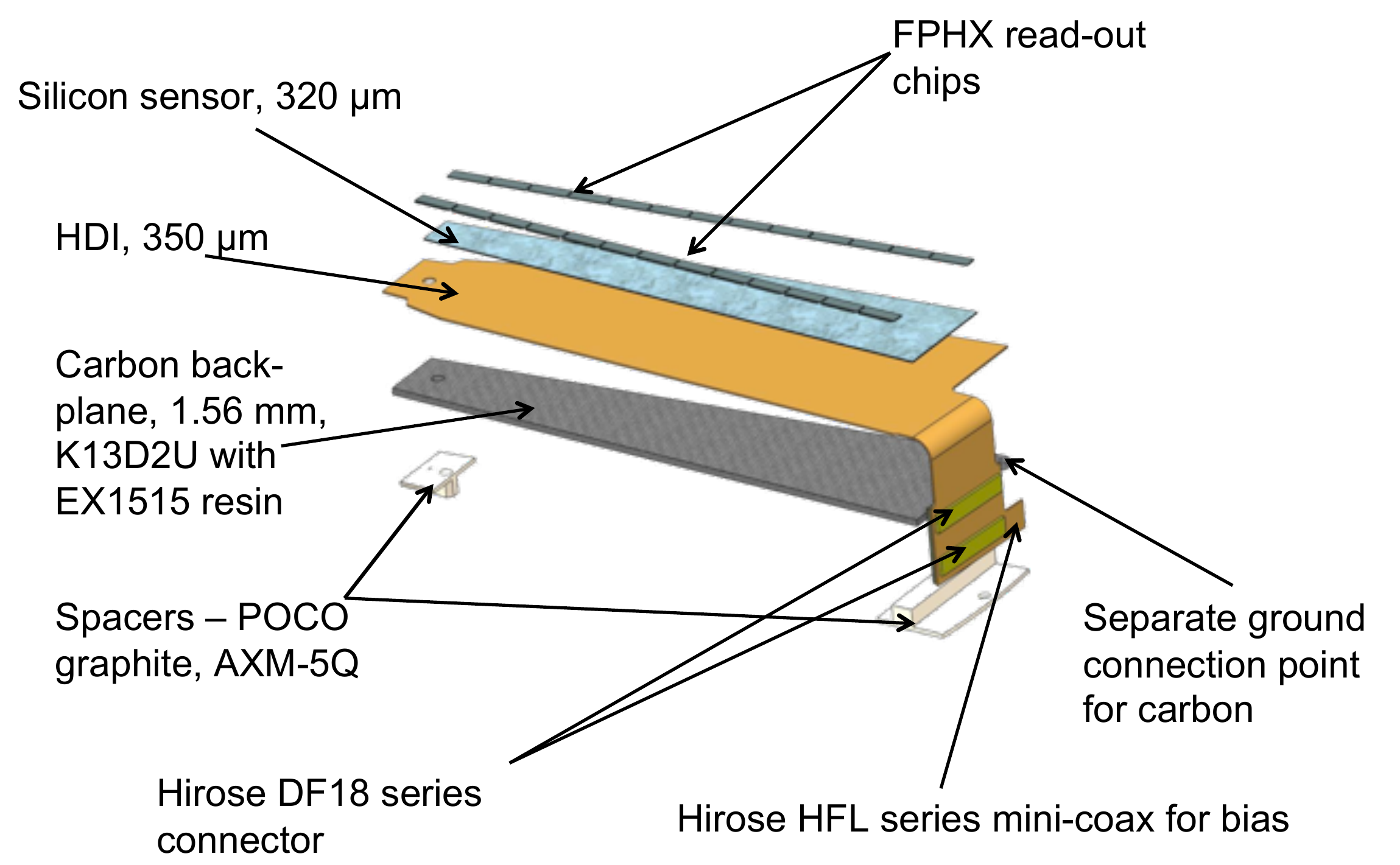}
  \caption{(color online) Exploded view of an FVTX sensor assembly.}
  \label{fig:sensor}
\end{figure}
A \emph{wedge} is the basic construction unit of the FVTX detector. Fig. \ref{fig:sensor} shows an exploded 
view of a single wedge assembly, which consists of a silicon mini-strip sensor (Section \ref{sec:sensor}), 
FPHX read-out chips (Section \ref{sec:chip}), a high-density interconnect bus (HDI, Section \ref{sec:hdi}), 
and a carbon support backplane.

\begin{figure}[htbp]
  \centering
  \includegraphics[width=0.8\linewidth]{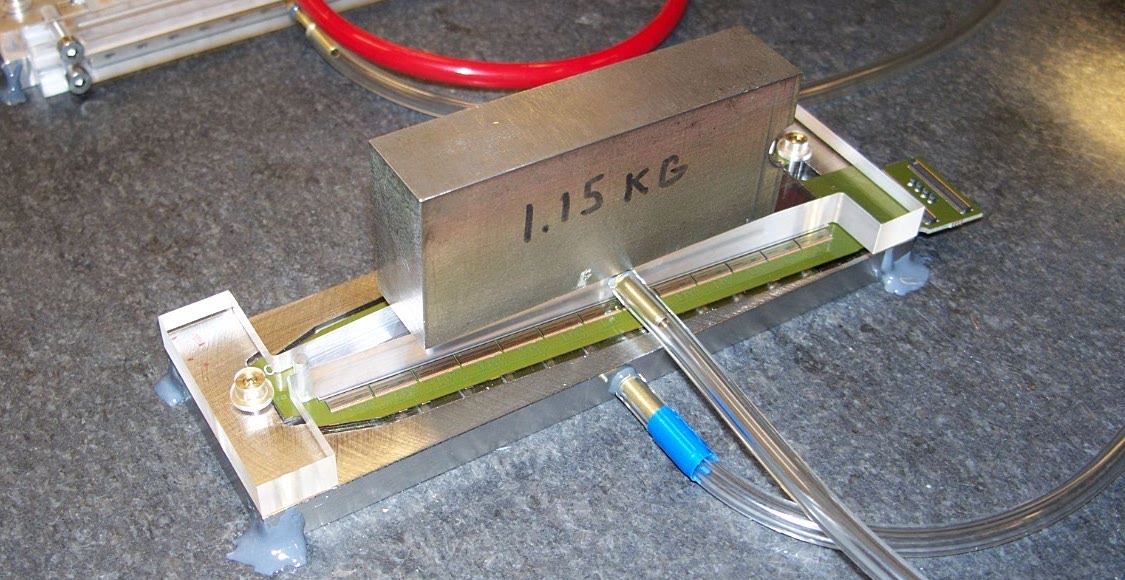}
  \caption{(color online) Assembly fixtures in use.  At this step, the silicon sensor is being placed onto the FPHX chip+HDI+backplane assembly.}
  \label{fig:wedge_jig}
\end{figure}

Assembly of wedges took place at the SiDet Facility at Fermilab. 
A series of precision assembly fixtures were used to affix the HDI onto the backplane, 
to place the FPHX read-out chips on the HDI+backplane assembly, and finally to attach 
the silicon sensor to the chip+HDI+backplane stack. At each assembly step, two fixtures used vacuum 
to hold the relevant components in place.  Pins and holes on the fixtures aligned the components as they were brought 
together, and pressure was applied between the two fixtures to ensure complete bonding between the components and adhesive. 
Fig. \ref{fig:wedge_jig} shows a wedge assembly in progress.

The carbon support backplane is formed from K13D2U carbon fiber (Mitsubishi Chemical) bonded with 
EX1515 (Tencate Advanced Composites) resin, with a total thickness of 1.56 mm.  Using the assembly fixtures, 
the HDI was glued onto the carbon backplane using Arclad 7876 \cite{Arclad}, a 50~$\mu$m-thick dry adhesive.  
The thin, seven-layer HDI circuit was populated with passive electronic components and connectors prior to mounting on the backplane.

After the HDI and backplane were assembled, a different set of assembly fixtures was used to place the FPHX chips in position on the HDI. 
Large wedges have 13 chips on each side, while small wedges have 5.  After placement the chips were wire bonded to the HDI and 
tests were performed to ensure proper function. A similar procedure was then used to attach the silicon sensor to the chip+HDI+backplane 
assembly with Arclad adhesive.  Silver epoxy (TRA-DUCT 2902 from Tra-Con) was used to connect the silicon's bias voltage surface 
to the HDI. Next, wire bonds were made between the sensor and the read-out chips.
A suite of tests were then performed on the completed unit to ensure proper functioning of the wedge.
After this the wire bonds were encapsulated with Dow Corning Sylgard 186 for protection. Completed wedge
assemblies were then shipped to Brookhaven National Laboratory for final preparation and installation.

Further operations were needed before a wedge was ready to be mounted
onto the disk support plane. A small ground wire was attached with conductive silver epoxy
to a small hole in the carbon plane, and connected to ground on the
HDI. This serves to drain off charge that may build up on the carbon
backplane. Next, graphite feet (Poco Graphite) were glued to the underside of the
carbon plane. Finally, the HDI was bent through $90^\circ$ near the
connector end using a thermal bending apparatus that heated the kapton-copper
HDI to 100$^\circ$C, in order to permanently form it to match the cage shape.

\subsection{Disks}
\label{subsec:disks}

\begin{figure}[htbp]
  \centering
  \includegraphics[width=0.8\linewidth]{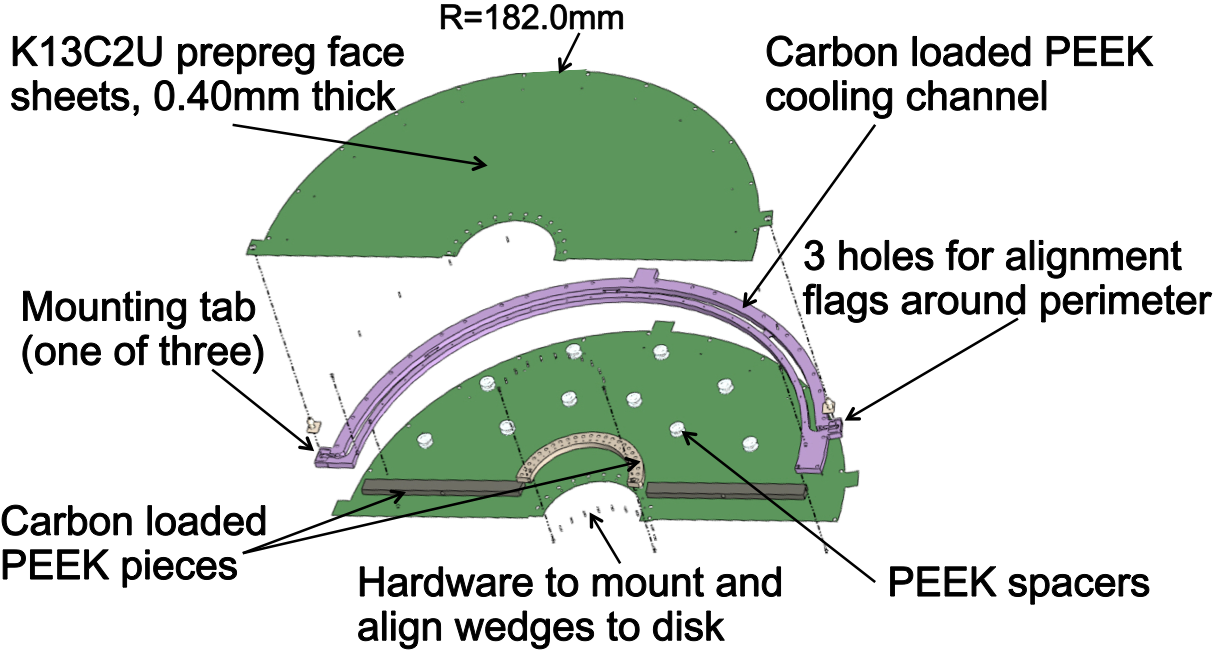}
  \caption{(color online) Exploded view of a support disk.}
  \label{fig:disk_exploded}
\end{figure}

The wedge support disks are flat sheets of 0.4 mm thick thermally conductive 
carbon fiber (K13C2U from Mitsubishi Chemical) on both sides of 
a carbon-loaded PEEK plastic frame.  The PEEK at the outer
radius contains a cooling channel, with nylon hose barb fittings at
the ends, which removes heat generated by the FPHX chips. 
PEEK buttons maintain the spacing between the face sheets.
Fig. \ref{fig:disk_exploded} shows an exploded view of a disk assembly. 

On both faces of the disks, precision alignment pins are located along the inner
and outer radii, one pair for each wedge which will be mounted on the disk. These pins match a precision
hole and slot in the graphite feet of the wedges, assuring placement
of the wedges onto the support disk with an accuracy of
25~$\mu$m. Wedges are secured onto the disk with small PEEK screws
near each of the alignment pins.

\begin{figure}[htbp]
  \centering
  \includegraphics[width=0.8\linewidth]{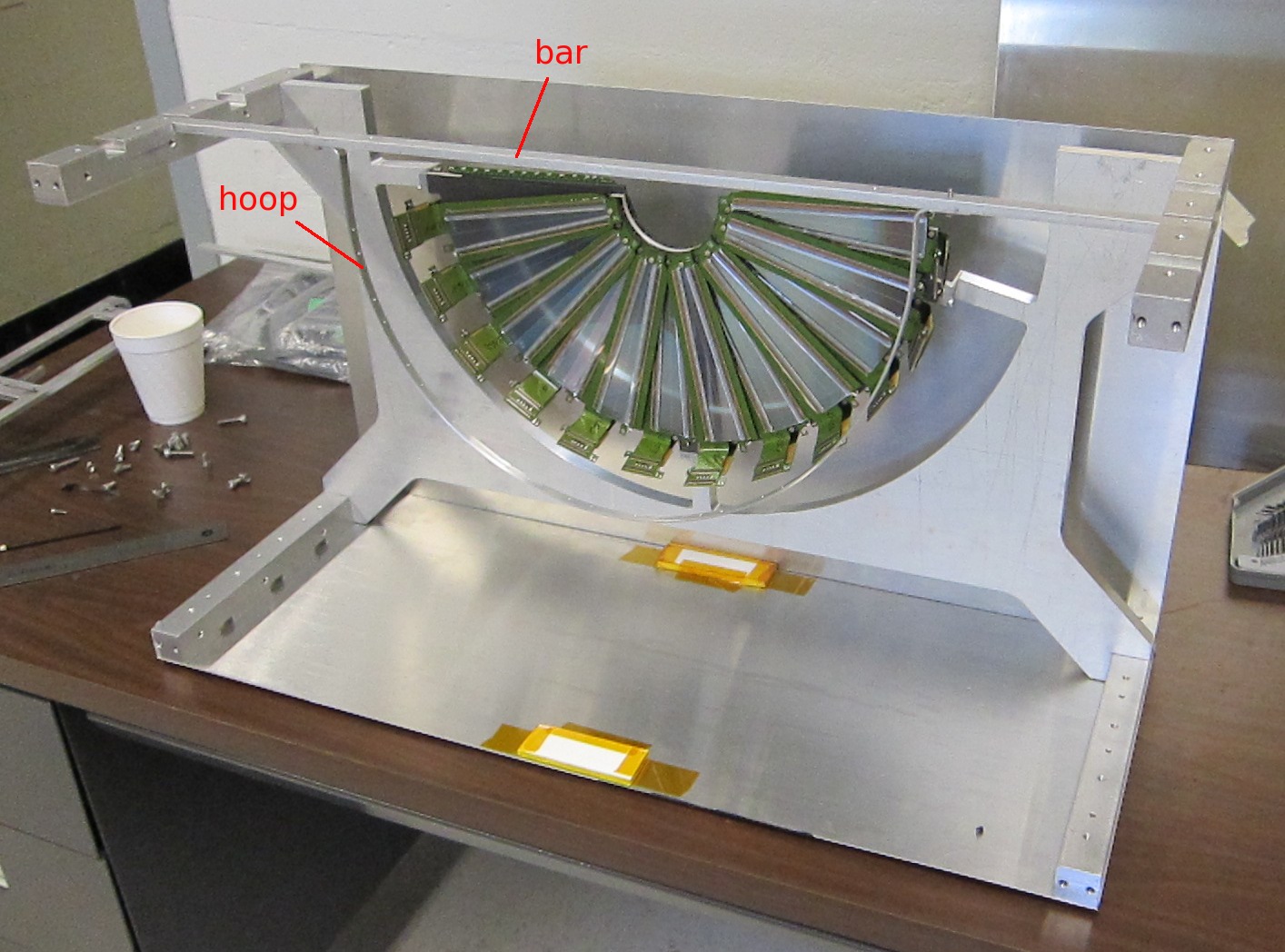}
  \caption{(color online) A populated disk in its support frame. Some of the cover sheets
are in place. The hoop will support the extension cables that will be 
connected to the wedges prior to installation in the cage.}
  \label{fig:disk_in_box2}
\end{figure}

The silicon sensor on each wedge subtends an angle of 7.5$^\circ$. Adjacent
wedges on a disk overlap in the azimuthal direction by 0.5 mm to give
hermetic coverage in the azimuthal direction.
Since the HDI is significantly wider than the silicon sensor, the wedges must be staggered in $z$ 
to allow this continuous azimuthal coverage by the sensors. This is achieved by mounting wedges on both sides 
of the disks, at alternating $z$ positions on each side.
The graphite feet on the back of the wedges come in two
varieties so that wedges can be alternately mounted at 0.9 mm or
4.9 mm above the surface of the disk.  During assembly, the disk was mounted on the precision mount 
points into an assembly frame. The frame allowed one to work on the
disk assembly in any orientation. In addition, aluminum cover sheets
could be mounted on the outside of the assembly frame, turning it into a
storage and transportation box, as shown in Fig. \ref{fig:disk_in_box2}.

After each disk was fully populated with wedges, the positions of the
wedges were precisely measured by Hexagon Metrology using an optical coordinate measuring
machine. Fiducial marks, four on each silicon
detector, were measured with an accuracy of 5~$\mu$m in the plane of the
silicon, relative to the three precision mounting points on the
perimeter of the disk. 

Before a disk was mounted into a cage, a temporary transfer fixture was
fitted to the centerline of the disk. Extension cables (section
\ref{sec:extension}) were connected to the wedges, as well as bias
voltage coaxial cables and disk cooling tubes.  The transfer fixture also
supported these cables. At this point, the disk was disconnected from
its three precision mount points on the assembly frame and was
transferred into a cage.

The disks were mounted into
a cage on the three precision mount points located on the outer
radius of the disk.  To maximize the detector's $\phi$ resolution, each of the four disks
are mounted into the cage offset in $\phi$ by an angle 3.75$^\circ$/4 with respect to the neighboring disk.

\subsection{Cages}
\label{subsec:cage}
\begin{figure}
  \centering
    \includegraphics[width=0.5\linewidth]{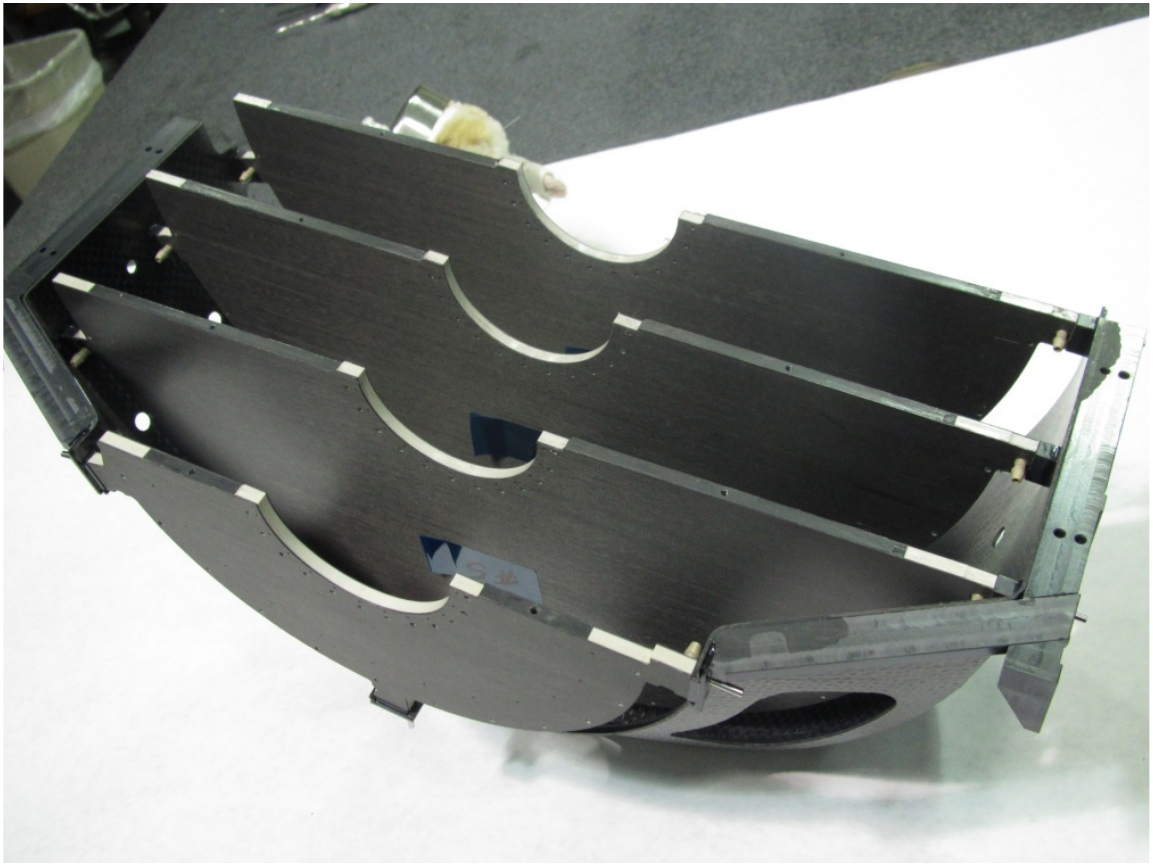}
    \caption{(color online) A cage with all four disks installed.  No wedges have been placed on the disks.}
  \label{fig:support}
\end{figure}

\begin{figure}[htbp]
  \centering
  \includegraphics[width=0.9\linewidth]{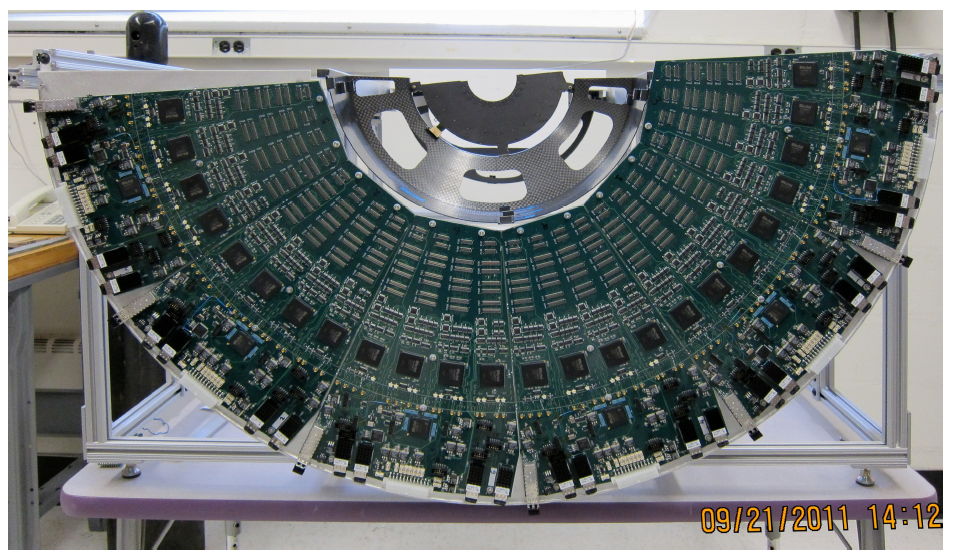}
  \caption{(color online) A cage and ROC boards on an assembly frame. An unpopulated small disk is mounted at the rear of the cage.}
  \label{fig:cage_roc}
\end{figure}

\begin{figure}[hbtp]
  \centering
    \includegraphics[width=0.9\linewidth]{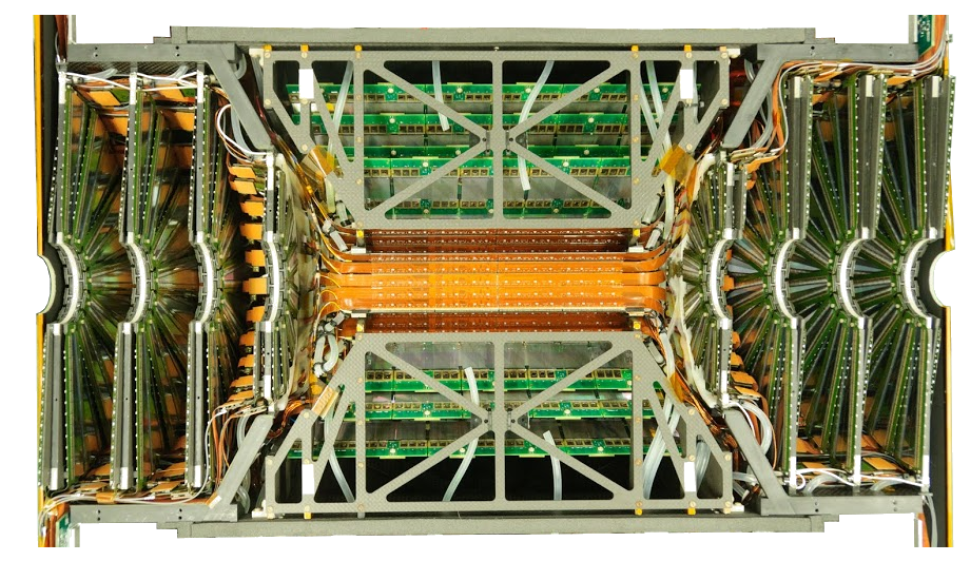}
    \caption{(color online) A completed half-detector, with the VTX barrels in the center,
and the two FVTX endcaps on either end. The overall length is 80 cm.}
  \label{fig:half_detector}
\end{figure}

The cages, into which disk assemblies
are mounted, are carbon-composite structures fabricated from CN60 carbon fabric (Nippon Graphite Fiber)
with EX1515 resin.  One of these cages is shown in Fig. \ref{fig:support} with four mounted disks (without wedges).  
During construction, the cage was mounted in an assembly structure that also
supported the aluminum cooling plate onto which the ROC boards (6 per quadrant) are
mounted, as shown in Fig. \ref{fig:cage_roc}. 
A soft material (Gap-Pad by the Bergquist Company),
approximately 1/8 in thick, is placed between the ROC and the cooling plate to improve heat transfer. 
Each disk is mounted into a cage on three mount points, each of which has an alignment pin and a screw. 
First the small disk was mounted, and extension
cables connected to the ROC boards, followed by the three large disks
in turn. At the inner radius of
the ROC boards, pairs of connectors can be seen, one pair for each
wedge/extension cable. A completed half-detector is shown in Figure \ref{fig:half_detector}.

During the selection for the support materials several factors were considered: 
low radiation length is desirable to minimize interactions in 
detector materials; high rigidity is necessary for maintaining alignment and stability of detector
components; ease of machining and availability are important for construction. 
Candidate materials for the FVTX support structures were beryllium, glass 
fiber reinforced polymer, and carbon-carbon composite. 

The wedge backplanes were constructed from K13D2U prepreg with EX1515 
resin, 1.56 mm thick. This fiber was selected because of its good thermal conductivity, which is necessary to remove heat 
generated by the FPHX read-out chips.
For the 0.4 mm thick disk face sheets, K13C2U prepreg with EX1515 resin was chosen because 
it is widely available, works well in sandwich
composites, and has a small radiation length and favorable strength
properties. 
For similar reasons, the cage was made from 6 plies of CN-60 cloth, for a thickness
of 1.5 mm.

The vibrational mode frequencies, gravitational load distortions, and shape
changes with temperature were studied for all mechanical structures,
and used to verify that the dimensional stability requirements were met.

\subsection{Cooling}
\label{subsec:cooling}
Heat generated by the FPHX chips on the wedge assemblies amounts to
390~$\mu$W/channel, or 1.3 W for a large wedge. The heat is conducted from the chips to the HDI, 
through the wedge carbon backplane, through the graphite
feet to the skin of the support disks, and finally to the cooling
channel at the outer radius of the disks.
With the cooling channel located at the outer radius of the wedge, a temperature gradient of 
$\sim$4$^\circ$C is developed across the length of the wedge during normal operation.

The ROC cards are mounted to a 1/8 in aluminum plate, with approximately 1/8 in of Gap-Pad
in between. These plates are cooled via chilled fluid 
circulation through an aluminum tube dip-brazed to the outer
edge of the plate.

\begin{figure}[htbp]
  \centering
 \hspace*{-2cm} \includegraphics[width=1.2\linewidth]{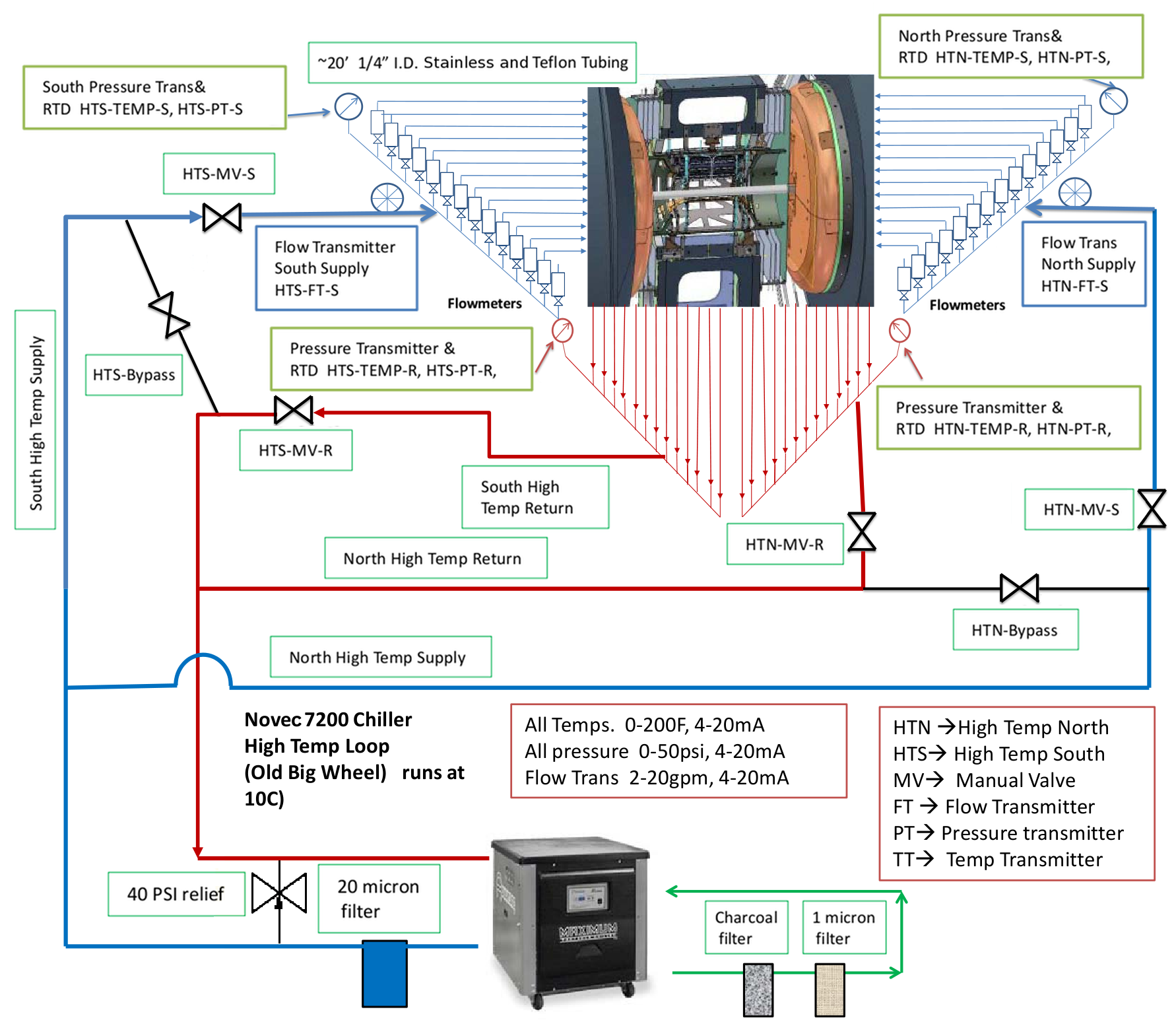}
  \caption{(color online) Schematic of one FVTX cooling loop.}
  \label{fig:cooling}
\end{figure}

The cooling system for the FVTX consists of two identical closed loops
operated at different temperatures.  One
loop provides 0$^\circ$C cooling fluid to the disks that the wedges are
mounted on, while the other provides 10$^\circ$C fluid to the aluminum cooling plates 
holding the ROC boards.  The systems are comprised of a chiller (stainless wetted
parts), stainless steel transfer lines, and a manifold and flow
control system, with short PTFE and Tygon tubing section where flexibility is required.  
The system also has a continuously running parallel
cleaning loop to remove contaminants.  The detector and manifold
system are located inside our interaction area and are inaccessible
while the collider is operating.  Because of this, a remote monitoring
and interlock system was developed to prevent detector damage in the
case of accidental loss of cooling.  A schematic of one of the systems
is shown in Fig. \ref{fig:cooling}.

The coolant used in the system is Novec 7200 (ethoxy-nonafluorobutane)
manufactured by 3M \cite{Novec}.  Novec 7200 is an engineered fluid that was
developed as both a heat transfer fluid and a cleaning fluid.  Novec
7200 is clear, non-conductive, non-corrosive, low odor, and a low
toxicity fluid that has zero ozone-depletion potential.  In addition,
it has very low greenhouse gas properties, and can be used as a heat
transfer fluid in the temperature range -138$^\circ$C to 76$^\circ$C.
The coolant leaves no residue behind when it evaporates.  These
characteristics make it very attractive from an environmental
and performance standpoint, but introduce several challenges when
using it as a cooling fluid.

Because of its cleaning ability, both material compatibility and fluid
cleanliness are concerns.  Novec 7200 has a low viscosity and low
surface tension. As a result, the fluid will work its way into tight
areas and dissolve any surface contaminants.  It will also dissolve
many plasticizers and components of soft materials, so proper
selection of wetted materials should be made.  Any contaminants left
in the fluid can then leave a residue on a surface in the event of a
system leak. Because of this, it was found that the fluid must be
cleaned while in operation to prevent contaminant buildup.  This is
achieved by continuously routing a small portion of the coolant 
through a charcoal column.

Another concern with the fluid is its electronegativity and
triboelectric charging while flowing through a non-conductive
material.  Problems arose where static charge built up in
Tygon and PTFE tubes.  The resulting discharges created pinhole
leaks through 1/32 in PTFE tube walls.  This can be avoided by using only
conductive tubing or minimized by reducing the length of
non-conductive tubing.

\begin{figure}[htbp]
  \centering
  \includegraphics[width=1.0\linewidth]{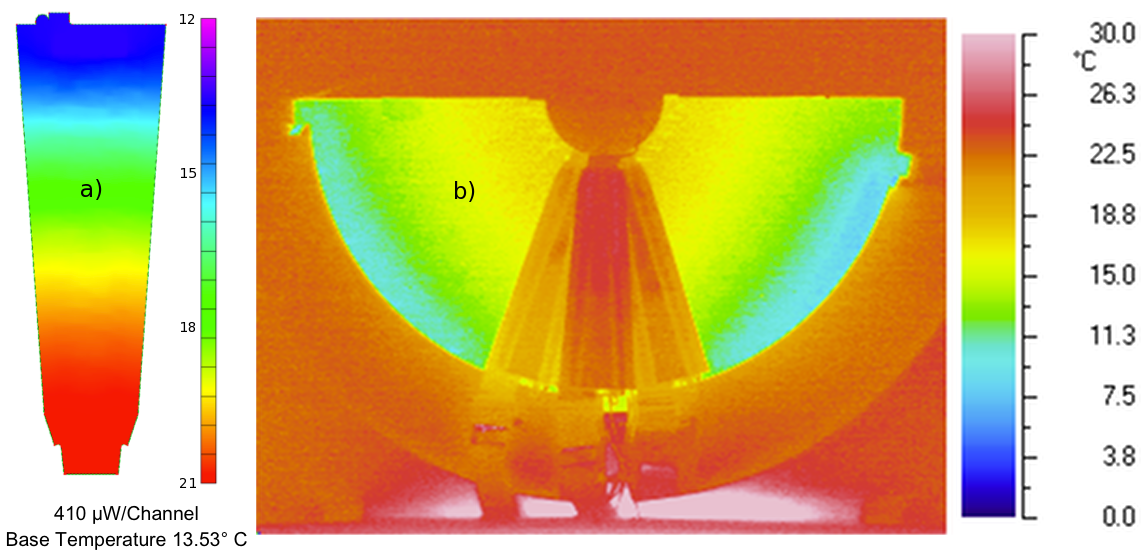}
  \caption{(color online) Calculated a) and measured b) temperature distribution of a wedge, powered and mounted on a disk, with 
cooling turned on.}
  \label{fig:wedge_temp_both}
\end{figure}

At the design stage, finite element calculations were performed to estimate
the temperatures and temperature gradients in the wedges. Fig.  \ref{fig:wedge_temp_both}
shows the expected temperatures on a wedge in the left panel. Since the cooling channel runs along the outer diameter
of the support disk, the tip of the wedge is warmer than the broader end. The gradient in this
calculation spans about 7$^\circ$C. The right panel shows an image taken with a thermal camera of a disk with three wedges mounted, while the wedges are powered and the cooling is running. 
The measured temperature gradient in the wedges is smaller than in the calculation, about 4$^\circ$C.

\subsection{Environmental Enclosure}  
\label{subsec:enclosure}
The central rapidity silicon vertex detector (VTX) and the two FVTX
endcap regions share an environmental enclosure, which is continually
purged with dry nitrogen gas during operation to avoid condensation on
the cooled electronics. The dry nitrogen is produced by bleeding the boil-off 
of a large liquid nitrogen dewar through a heat exchanger to raise the temperature
of the gas. The enclosure radius is 22 cm in the region
of the VTX barrel, and extends to $\sim$50 cm from the beam axis 
to contain the VTX transition electronics boards and the FVTX ROC boards.

\section{Radiation Effects}
\label{sec:radiation}

\begin{figure}[htbp]
	\centering
	\includegraphics[trim=0cm 2.2cm 0cm 0cm, clip=true, width=1.0\linewidth]{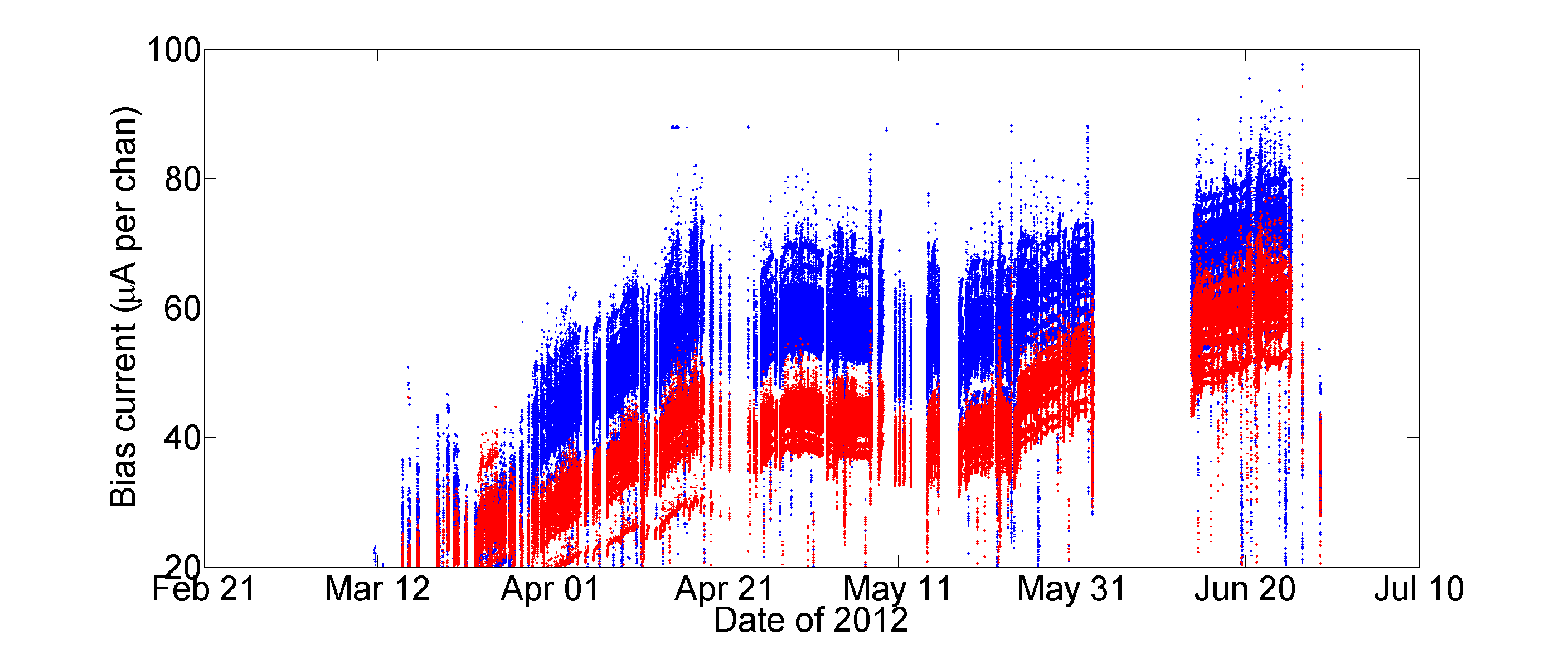}
  \caption{(color online) Current draw as a function of time for the FVTX bias supplies during the 2012 RHIC run. Each point is 
from the bias supply for a group of 8 individual wedges.  The North arm is in blue and the South arm is in red.}
	\label{fig:bias}
\end{figure}

Silicon detectors operated in high-radiation environments at colliders
can suffer from radiation damage to the bulk silicon,
resulting in increased leakage currents and noise
\cite{CDFsilicon,D0silicon,silicon_review}.  The FVTX location near
the collision vertex exposes the silicon sensors to a
significant ionizing radiation flux during beam tuning and data
taking.  In the 2012 run at RHIC, the FVTX was exposed to radiation
from Cu+Au, $p+p$, and U+U collisions at $\sqrt{s_{NN}}$ = 200 GeV,
and $p+p$ collisions at $\sqrt{s}$ = 510 GeV.  As expected,
significant increases in the wedge leakage currents were observed. 
Fig. \ref{fig:bias} shows the current draw from the FVTX bias power 
supplies as a function of time during the 2012 run.

In addition to direct radiation produced by colliding beams, secondary radiation
induced in material surrounding the FVTX may also contribute to the
bulk radiation damage.  Specifically, the nosecone pole tip of the PHENIX central magnet, 
located directly behind the FVTX, is a possible source of secondary electrons, x-rays, and spallation neutrons.

To study the effects of radiation in the actual operating environment
of the FVTX, active radiation-monitoring sensors were installed
in the PHENIX interaction region during the RHIC run starting in
February 2013, which was dedicated to $p+p$ collisions at $\sqrt{s}$ =
510 GeV.  The sensors were procured from the Solid-State Radiation
Sensor Working Group at CERN.  Each PCB carrier (15 mm $\times$ 32 mm) was
populated with a p-i-n diode, a RadFET, and a temperature probe. The device was monitored by
providing a constant current and reading out the required voltage.
Sensors were mounted on the nosecone
pole tip of the PHENIX central magnet at radial distances (r = 3.5, 8.5
and 16.2 cm) corresponding to the inner and outer radii of silicon on
the FVTX station disks.  During the initial beam tuning,
when the VTX/FVTX was positioned several cm away from the final
position around the beam pipe,
two additional sensors were mounted at distances
along the beam axis corresponding to the locations of the innermost and outermost FVTX 
station disks (see Table \ref{tab:mech_summary}). 
These sensors were removed when the FVTX/VTX assembly was closed from the
retracted to the operating configuration.  A BNL-designed front-end
electronics board capable of interfacing with all the sensors
facilitated real-time monitoring through the current PHENIX online
system.  Finally, five passive thermoluminescent dosimeters (TLDs)
were mounted at positions corresponding to the CERN sensors to cross
check the integrated dose. These were removed when the detector was
moved into its operating position.

\begin{figure}[htbp]
	\centering
\includegraphics[trim=0cm 0cm 0cm 1.05cm, clip=true, width=0.9\linewidth]{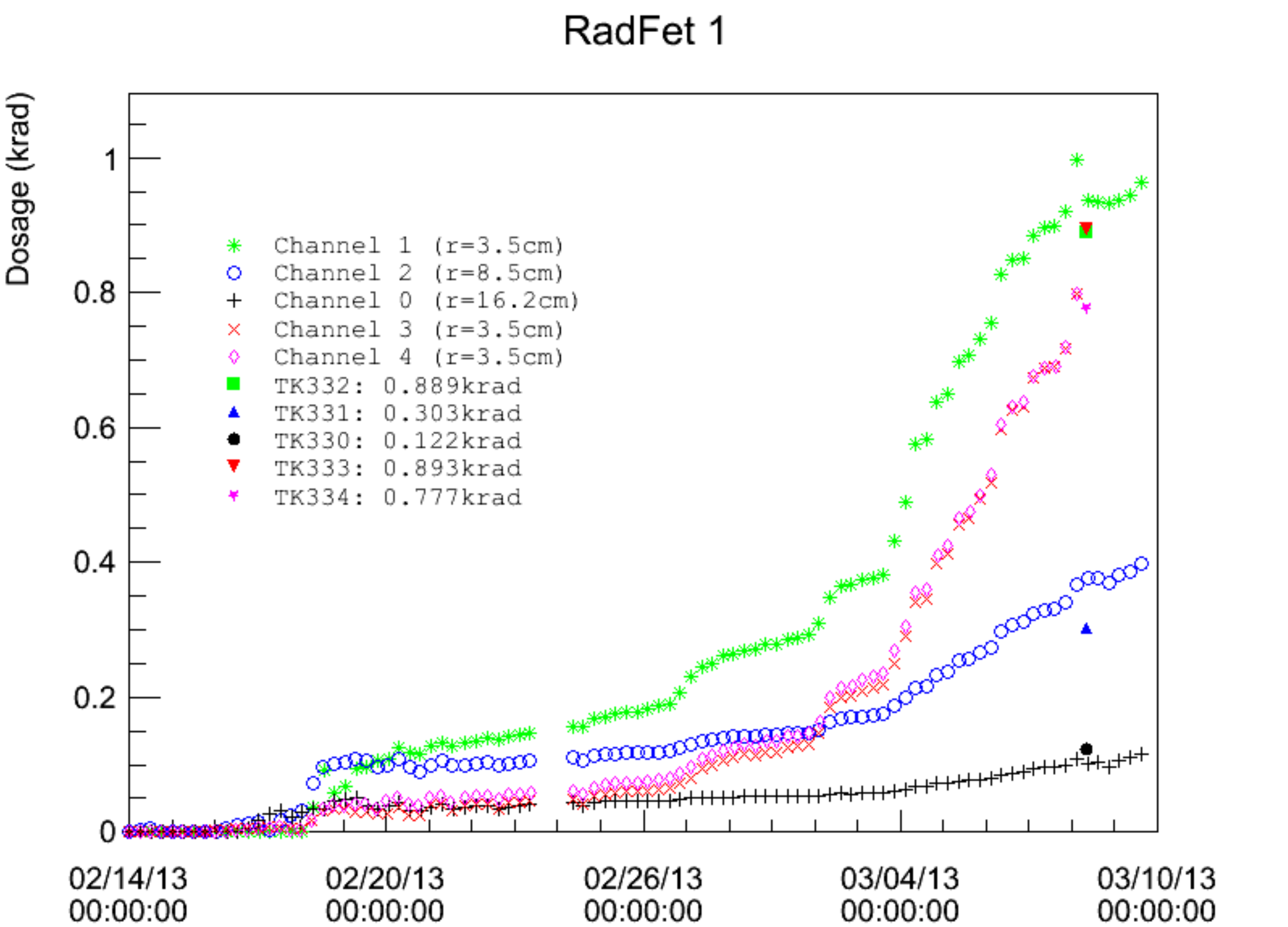}
\caption{(color online) Integrated dose in the RadFET monitors as a function of time
during RHIC beam tuning for the 2013 run. The TLD dose is shown for reference.}
	\label{fig:RadFET_1}
\end{figure}

\begin{figure}[htbp]
	\centering
  \includegraphics[trim=0cm 0cm 0cm 1.05cm, clip=true, width=0.75\linewidth]{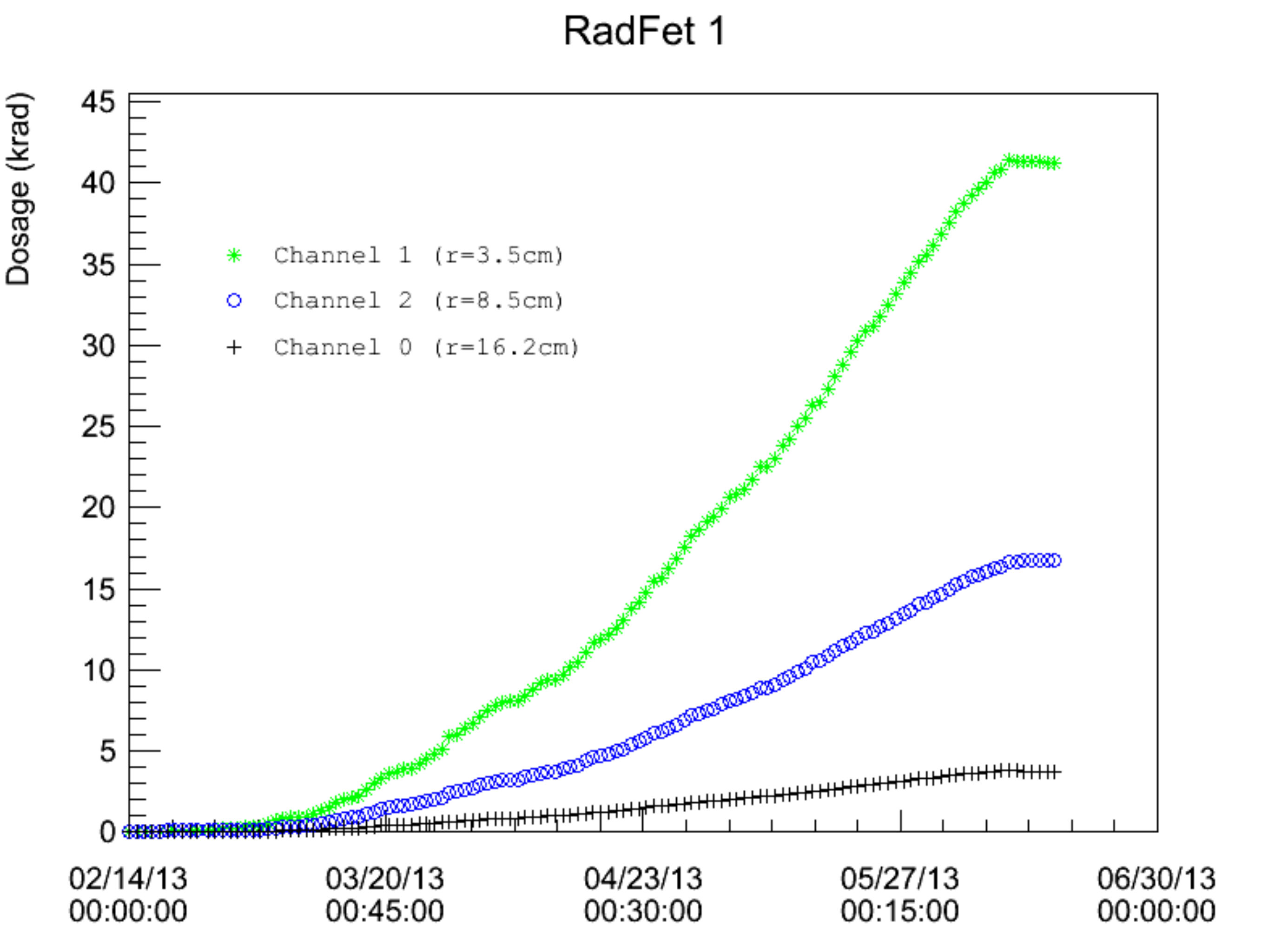}
  \caption{(color online) Integrated dose from the three RadFETs installed on the
PHENIX magnet poletip, directly behind the FVTX, measured as a function
of time during the 2013 RHIC run.}
  \label{fig:RadFET_2}
\end{figure}

Fig. \ref{fig:RadFET_1} shows the integrated dose measured during the
tune-up period of the collider, with the detector in the retracted
position.  Contributions to the integrated dose from individual stores
in RHIC are clearly visible.  The data exhibit the expected drop in
dose as the radial distance from the beam axis increases.  More
importantly, there is good agreement for the measured dose from gamma
radiation between the RadFETs calibrated with a method provided by
CERN and the TLDs (labeled TKXXX).  Fig. \ref{fig:RadFET_2} shows the
integrated dose measured for the entire 2013 RHIC run from 510 GeV $p+p$
collisions with the remaining three RadFETs on the nosecone
poletip.  The highest integrated dose is 43 krad for the RadFET (1
$\times$ 1 $\times$ 0.5 mm$^{3}$ ) at a radial distance of 3.5 cm from
the nominal beam axis.  This is well under the yearly dose that would 
limit operation of the FVTX to less than ten years (under
normal operating conditions at RHIC).  A similar conclusion can be
drawn for the integrated dose from neutrons measured with the p-i-n
diodes, which are systematically lower than the dose from gamma
radiation measured with the corresponding RadFETs shown in Fig. \ref{fig:RadFET_2}.

To quantitatively determine the effect of radiation damage on the
FVTX, several FVTX sensors were directly exposed to the 800 MeV proton
beam at LANSCE \cite{LANSCE}. Pairs of FVTX sensors were mounted in
FR4 frames and placed in the beam's path.  The dose
delivered during the 510 GeV $p+p$ portion of RHIC's Run-12 was taken
as a benchmark.  During that run, the PHENIX experiment sampled 30
pb$^{-1}$ within the nominal 30 cm vertex range.  Using measurements
of the radiation environment at RHIC from \cite{VTXrad}, this
luminosity corresponds to a delivered dose
of  $\sim$1.7$\times$10$^{10}$ N$_{eq}$/cm$^{2}$, where N$_{eq}$ is the
equivalent flux of 1 MeV neutrons.  Since the LANSCE beam has a
hardness $\kappa \sim 0.7$, this corresponds to a flux of 2.5$ \times
10^{10} p/$cm$^{2}$.  Pairs of wedges were exposed to 1, 5, 10, and 20
times this dose.  During irradiation, the sensors were at room
temperature and bias was not applied.  The leakage currents of the
wedges were measured $in$ $situ$ immediately preceding and following
beam exposure.  Prior to irradiation, all wedges drew less than 20 nA
at 100 V.

\begin{figure}[htbp]
	\centering
  \includegraphics[width=1.0\linewidth]{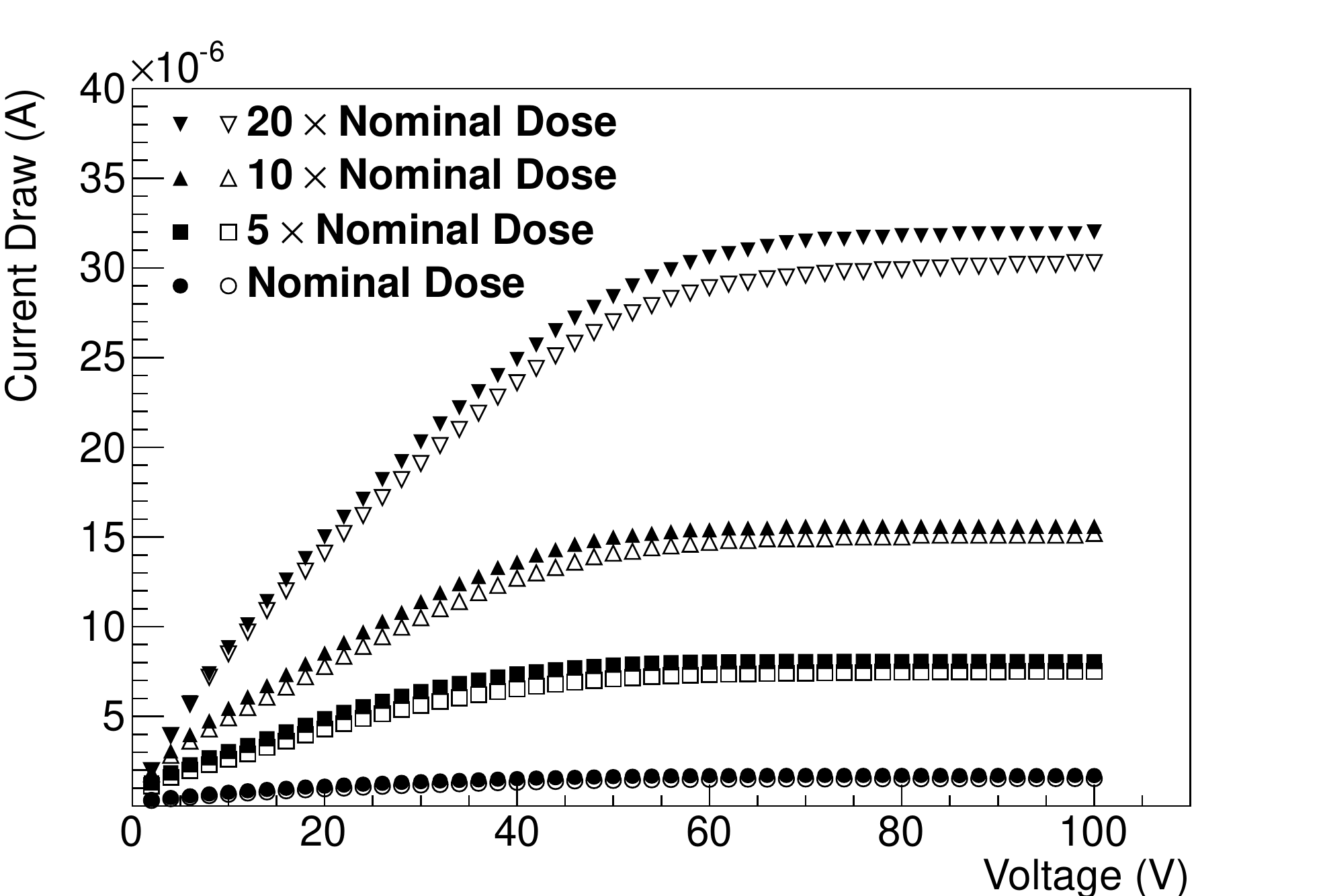}
  \caption{Leakage current as a function of the applied
    bias voltage for the test wedges, measured immediately following
    irradiation.}
	\label{fig:IV_1}
\end{figure}
 
\begin{figure}
  \centering
  \parbox{6cm}{
    \includegraphics[width=6cm]{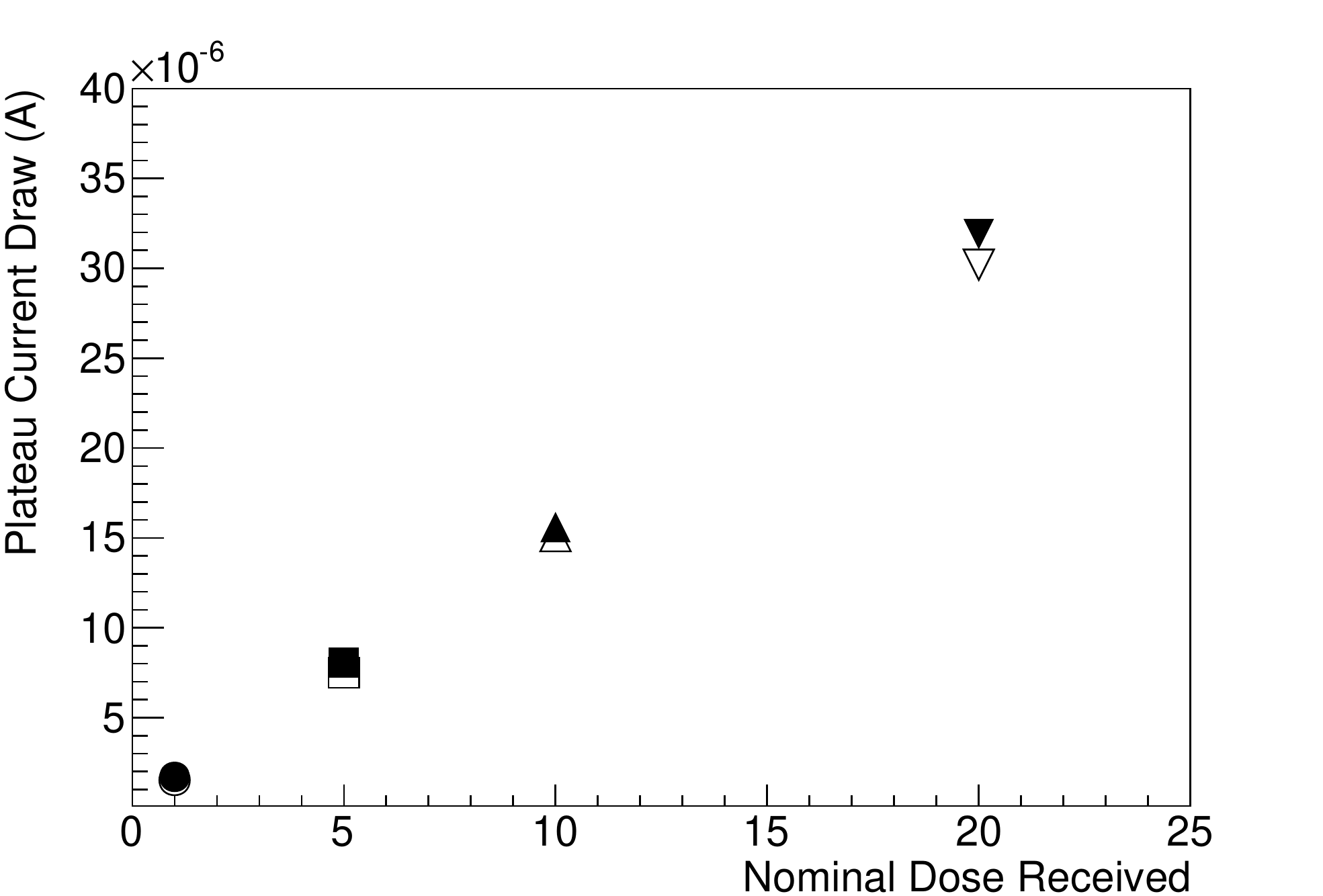}
    \caption{The plateau values of the leakage current of the
      irradiated wedges, measured immediately following the
      irradiation.}}
  \label{fig:the_plats_1}
  \qquad
  \begin{minipage}{6cm}
    \includegraphics[width=6cm]{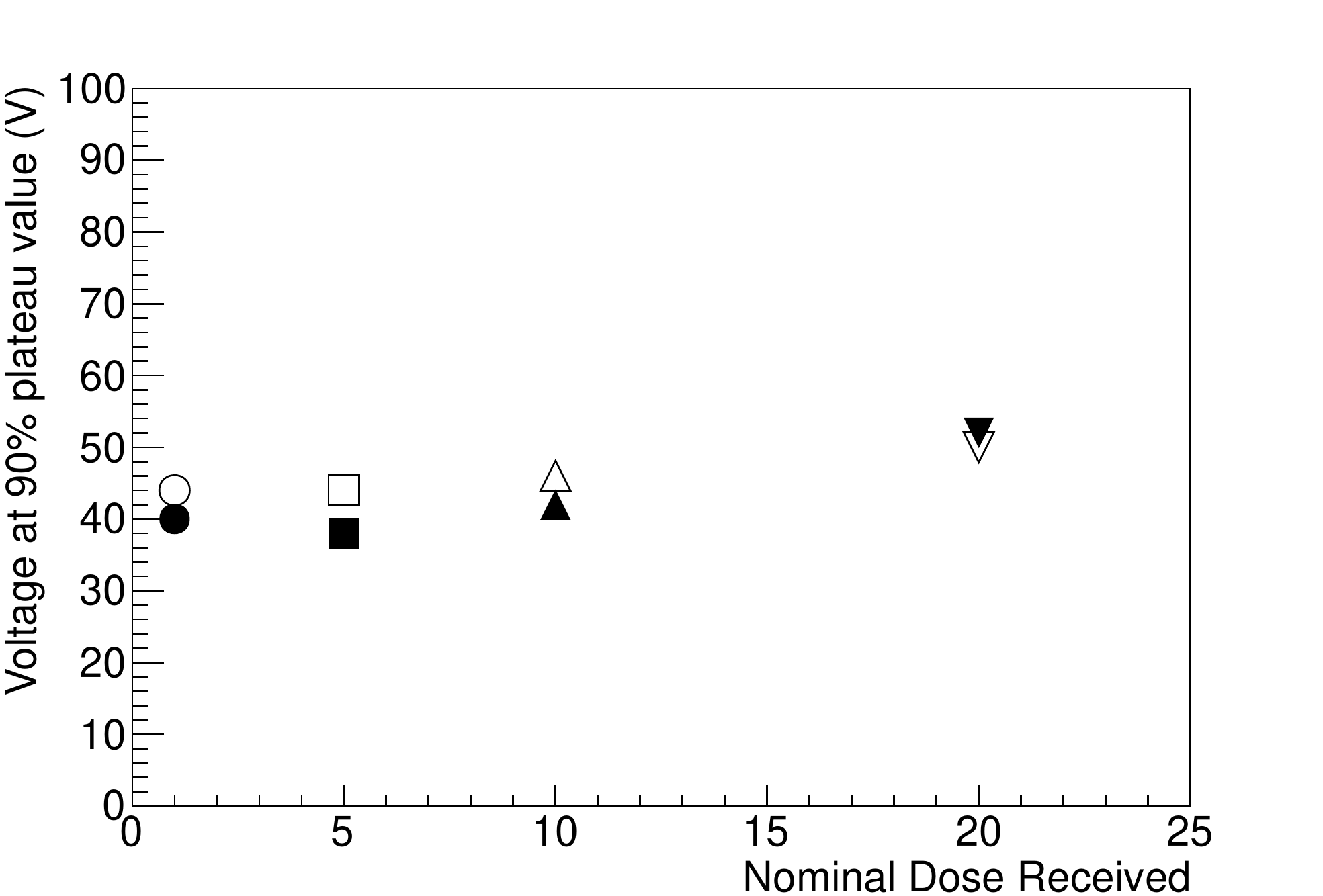}
    \caption{The applied bias at 90$\%$ of the plateau value of the
      leakage current, measured immediately following the
      irradiation.}
    \label{fig:voltages_1}
  \end{minipage}
\end{figure}

Fig. \ref{fig:IV_1} shows the leakage current as a function of bias
voltage for the sensors after exposure to the LANSCE proton beam.  As
expected, the leakage currents increased proportionally to the
received dose (see Fig. 27).  To examine the effects of
radiation on the bias voltage required for full depletion, the voltage
at which the leakage current is 90$\%$ of the plateau value is shown in
Fig \ref{fig:voltages_1}.  Although there may be a small increase in
the most irradiated sample, this is well below the typical operating
bias voltage of 70 V.  A small but detectable amount of residual
radiation ($<$ 1 mrem/hr on contact) was observed on the wedges after
irradiation, presumably due to activation of various isotopes in the
sensors and FPHX chip material.  This low amount of residual radiation
is not expected to significantly contribute to ongoing damage in the
bulk silicon.

It is well established that the electrical characteristics of
irradiated silicon vary with time and sensor temperature
\cite{6164289,256577,467878}.  In the short term, on the scale of
weeks to months, a decrease in leakage current due to annealing is
expected.  Long term effects due to reverse annealing, which happen on
the scale of years, can increase the leakage current and bias voltage required
for full depletion.

\begin{figure}[htbp]
	\centering
  \includegraphics[width=1.0\linewidth]{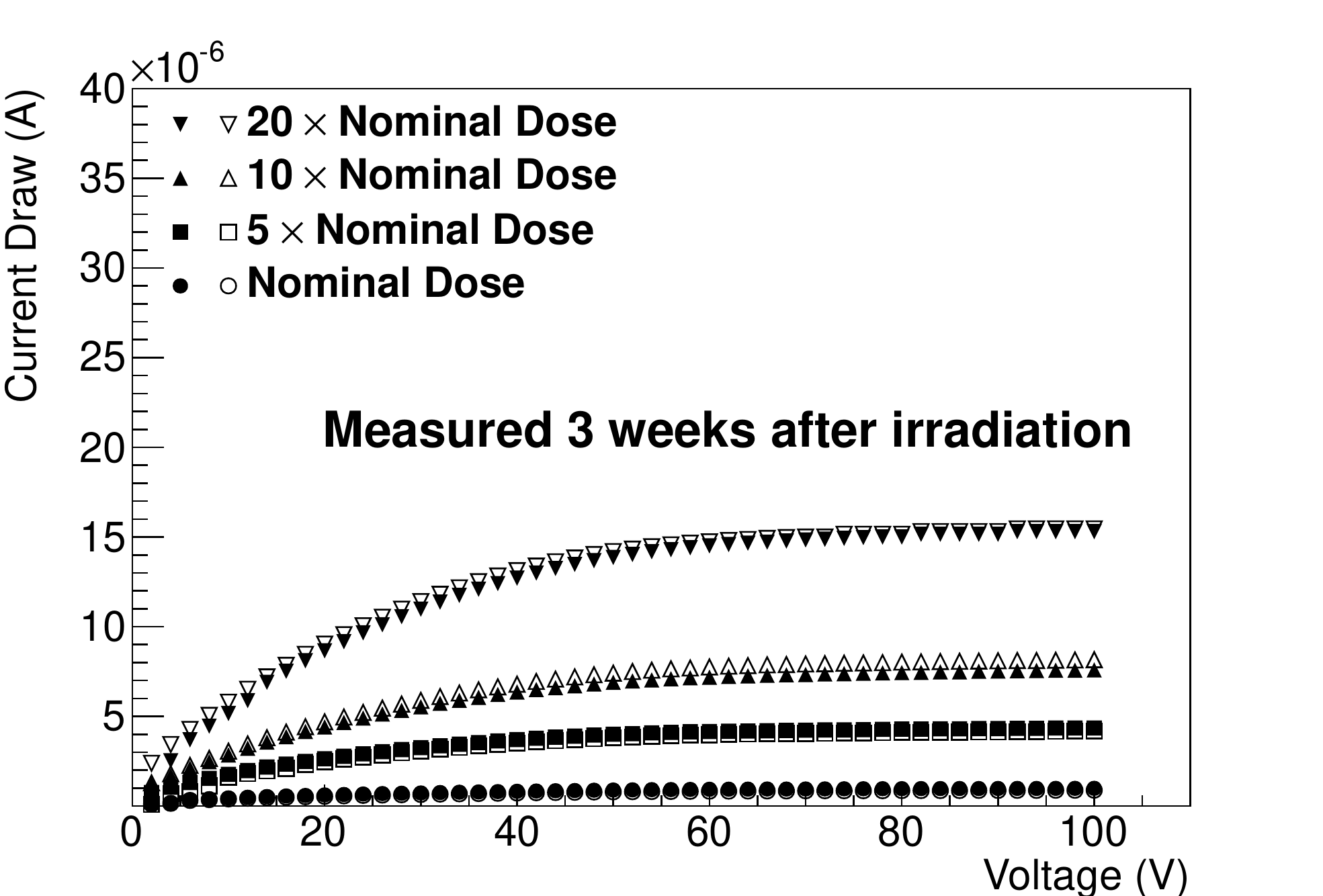}
  \caption{Leakage current as a function of the applied
    bias voltage for the test wedges, measured three weeks after the
    irradiation.}
	\label{fig:IV_2}
\end{figure}

\begin{figure}
  \centering
  \parbox{6cm}{
    \includegraphics[width=6cm]{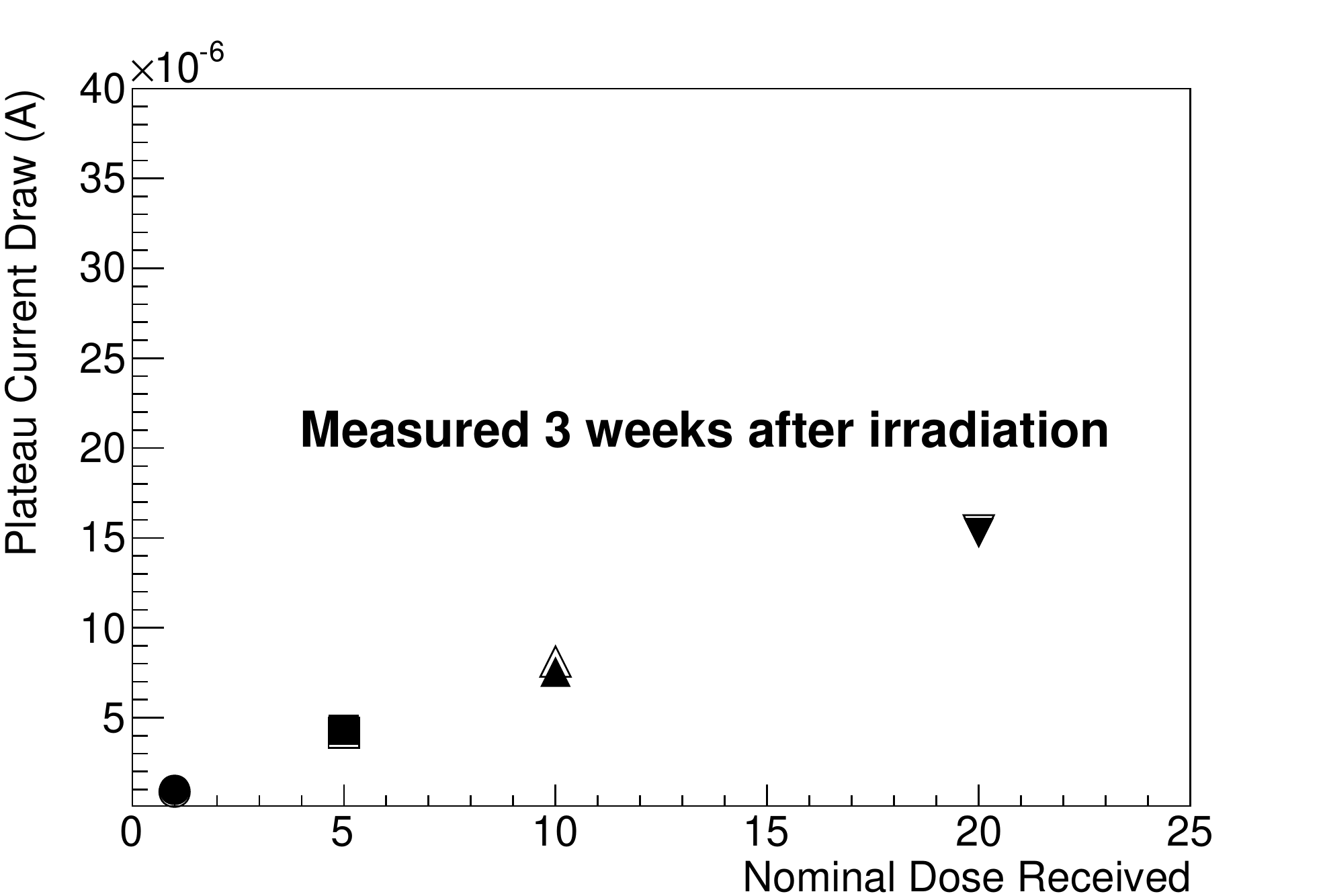}
    \caption{The plateau values of the leakage current of the
      irradiated wedges, measured three weeks after the irradiation.}}
  \label{fig:second}
  \qquad
  \begin{minipage}{6cm}
    \includegraphics[width=6cm]{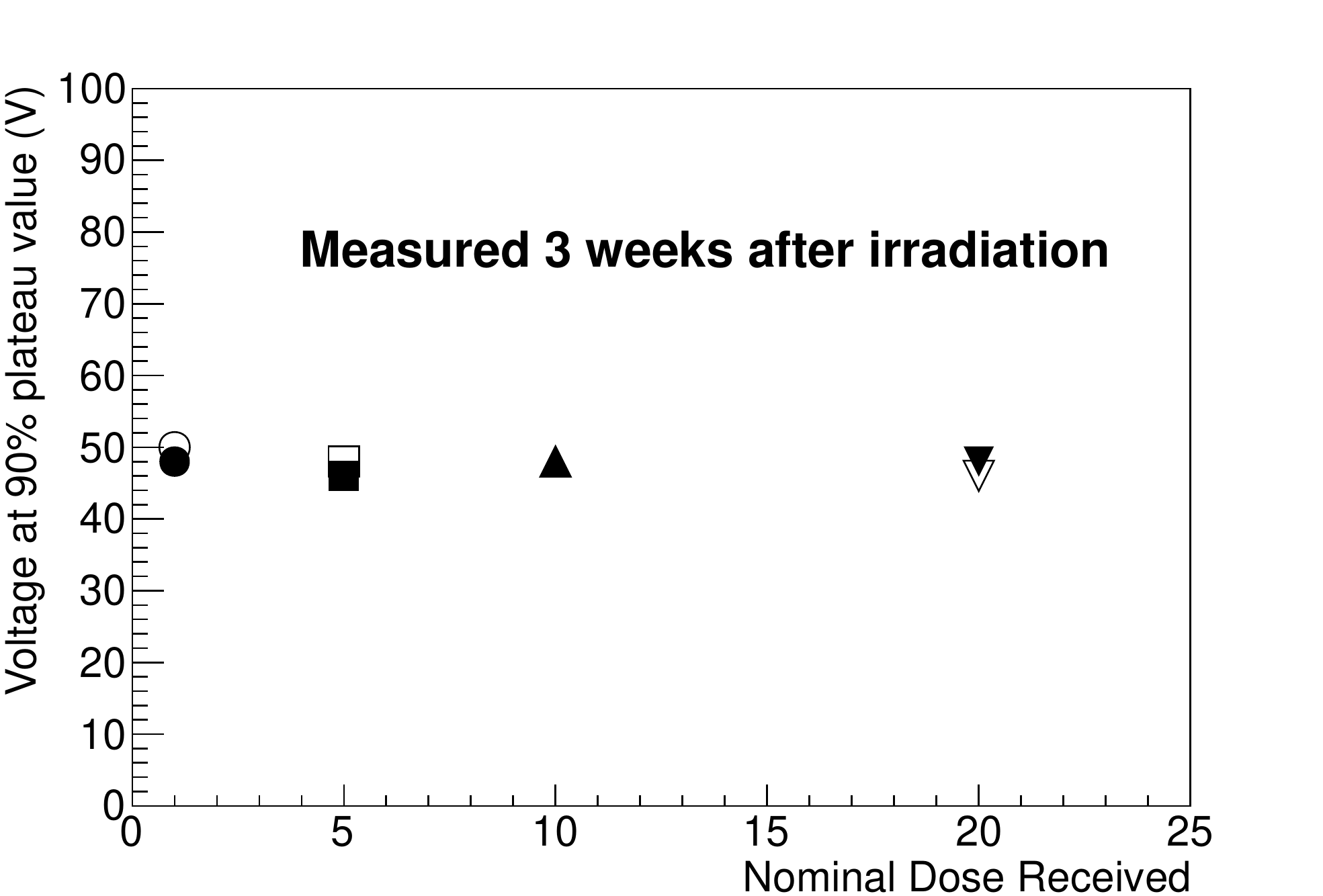}
    \caption{The applied bias at 90$\%$ of the plateau value of the
      leakage current, measured three weeks after the irradiation.}
    \label{fig:voltages_2}
  \end{minipage}
\end{figure}

To examine effects from annealing, the leakage currents were measured
again for each wedge after three
weeks.  Since the actual FVTX sensors are kept at room temperature in
between data taking periods, no temperature excursions were
undertaken, as these can affect the annealing properties of silicon.
Fig. \ref{fig:IV_2} shows the resulting I-V curves, after a correction
was applied to account for the slight difference in sensor temperature
at the time of the second measurement.  A significant decrease
($\sim40\%$ for each wedge) in the plateau leakage current is observed (see Fig. 30), which is
the expected effect of annealing.  No large changes in the required bias
to reach the plateau of the leakage current are observed (Fig. \ref{fig:voltages_2}).  Further studies of the evolution of
the leakage current over time are underway, in order to understand the
competing effects of annealing and any reverse annealing which may
occur.

While the FVTX sensors do show the expected increase in leakage
current with the received radiation dose, the magnitude of the
increases will not require any changes in the cooling system or bias
voltage equipment over the expected life of the experiment.
\section{Performance}
\label{sec:performance} 

This section presents performance benchmarks from the FVTX, using data
collected during operation at RHIC.

\subsection{Timing}

\begin{figure}[htbp]
	\centering
  \includegraphics[width=8cm]{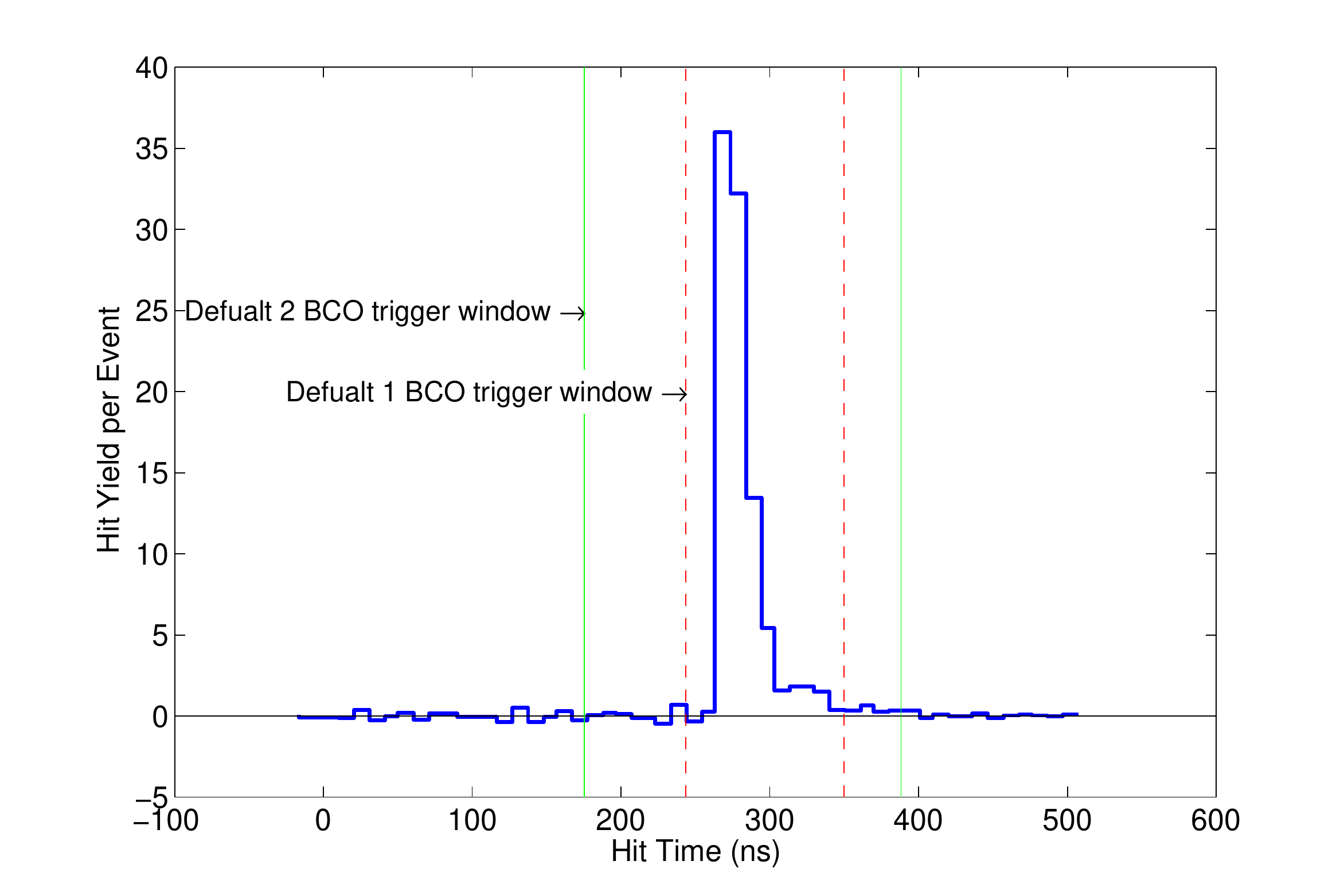}
  \caption{(color online) Timing distribution of the FVTX hits relative to the RHIC beam clock. }
	\label{fig:timing}
\end{figure}

The distribution in time of FVTX hits is studied relative to the RHIC collision
time by comparing the hit rate at different FVTX delay values relative to the RHIC beam
clock. The timing distribution for two sectors of wedges in
the south arm is shown in Fig. \ref{fig:timing}. Most hits fall in a window
$\sim$30~ns wide.

Two standard trigger timing configurations were used during FVTX operation, as
shown by the vertical lines in Fig. \ref{fig:timing}: during relatively low trigger rate running (in heavy ion systems)
hits arriving in a time window two RHIC beam clocks (BCO) wide
(1~BCO$\sim106$~ns) are accepted. In high trigger rate $p+p$ running, a 1 BCO-wide window
is used to avoid recording accidental hits from neighboring beam crossings (1 BCO apart).

\subsection{Hit Efficiency}

Multi-layer tracking detectors require a large intrinsic hit efficiency in each sensor,
that is, a high probability that a particle of interest will produce a measured signal when traversing an active
sensor layer.  To evaluate this efficiency in the FVTX, charged particle tracks which are identified by hits in
three of the FVTX stations are projected to the fourth station.  A hit cluster in the fourth station at the projected
position is assumed to be due to the charged track, which is a good assumption for the low occupancy $p+p$ events used
in this study.

\begin{figure}[htbp]
	\centering
  \includegraphics[width=8cm]{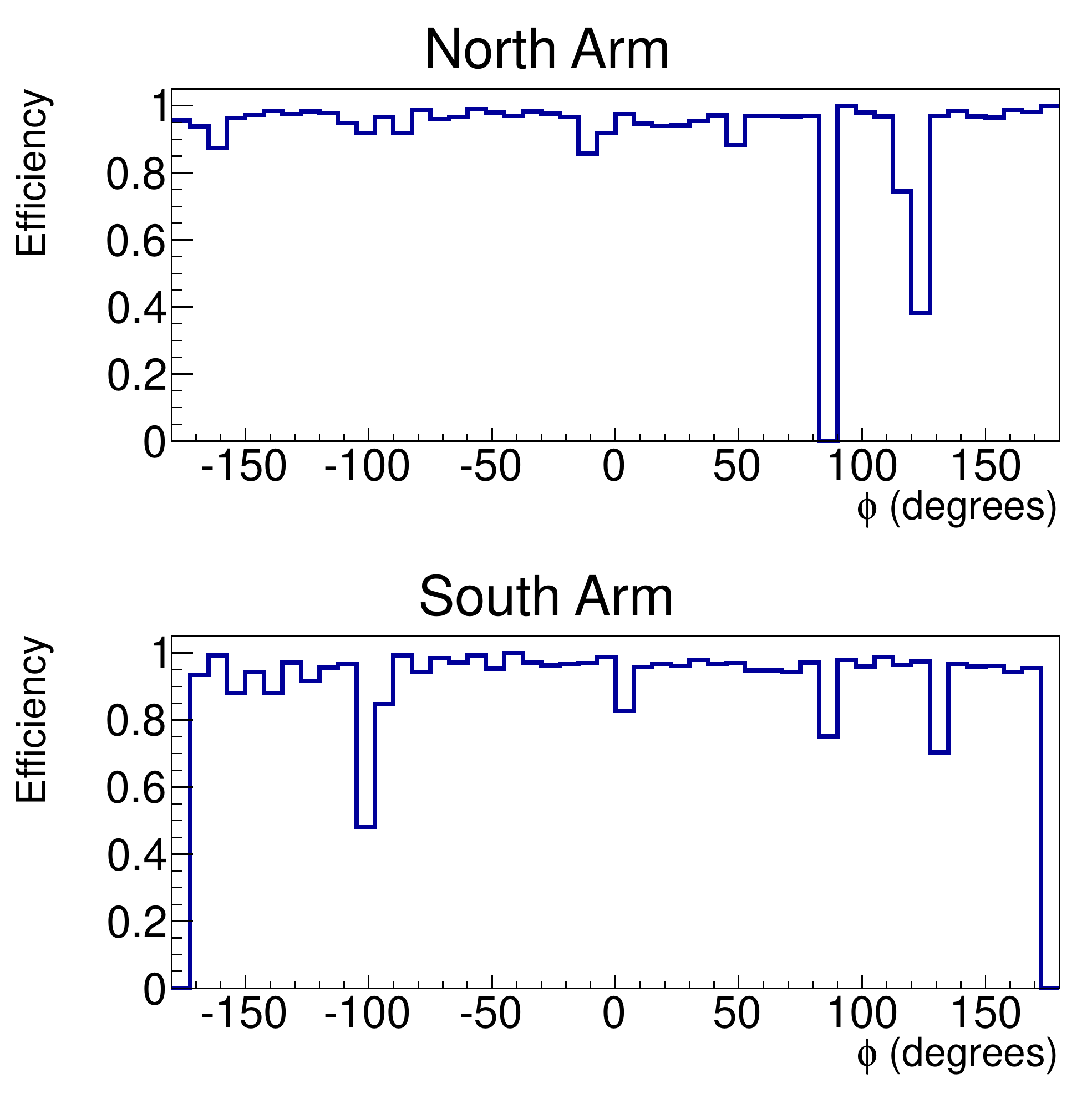}
  \caption{Hit efficiency for FVTX station 2 as a function of $\phi$. }
	\label{fig:FVTX_eff_phi}
\end{figure}

The probability of finding a hit at the projected spot in station 2 using tracks identified by hits in stations 0, 1, and 3 is shown in Fig. \ref{fig:FVTX_eff_phi},
as a function of the angle $\phi$ around the disk,
using data recorded during the 2013 RHIC $p+p$ run.
The extracted efficiencies shown in this plot include the intrinsic efficiency of the
detectors as well as any efficiency loss due to dead channels, chips, or
DAQ channels. The peak efficiencies are above 95$\%$ indicating that the intrinsic
efficiency of the detector is quite high.
The area near $\phi = 90^\circ$ in the North arm has a low hit efficiency due to a broken component on a ROC board,
which prevented several wedges from being read out.  However, the overall live area during the 2013
run was greater than 95$\%$.

\subsection{Alignment and Residuals}

Misalignment of the silicon wedges relative to each other and multiple scattering
of particles as they pass through the FVTX sensor material will have a detrimental
effect on the ultimate tracking resolution of the detector.  The internal detector
alignment was performed using data taken with the PHENIX magnets turned off, so all
charged particles travel in straight lines.  The MILLEPEDE-II~\cite{Blobel} package was
used to internally align all detector elements.

\begin{figure}[htbp]
	\centering
  \includegraphics[width=1.0\linewidth]{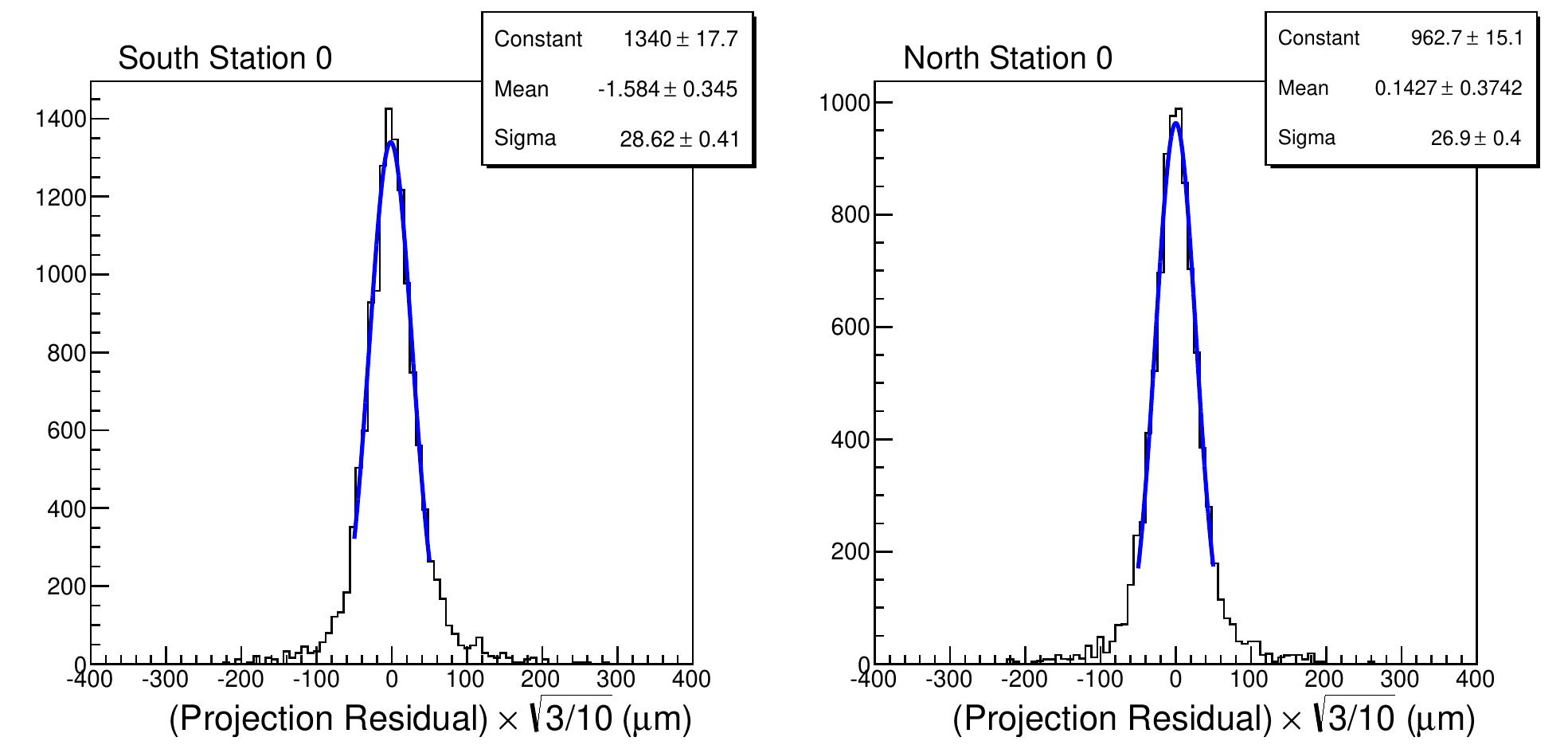}
  \caption{(color online) Track residuals for the innermost FVTX tracking stations, scaled to give the single hit resolution.}
	\label{fig:residuals}
\end{figure}

After detector alignment was performed, the FVTX single hit resolution was determined with straight-line tracks found in the FVTX, matched
with tracks found in the muon spectrometer, from $p+p$ collisions recorded with the PHENIX magnets turned off.  These tracks typically have 
a total momentum $p > 3$ GeV/$c$. After finding tracks with hits
in three FVTX stations, the track residual for the fourth station is found by calculating the distance between the track projection
and the center of the nearest FVTX hit cluster in that station.  The width of this track residual distribution is determined by the hit position resolution in each station 
and the distance between tracking layers.  To find the single-particle hit position resolution for a single station, a correction is applied to the track residuals, which was 
determined from linear regression assuming a common single-particle hit position resolution in the three stations used to find the track and a common distance between the stations.
The scaled track residuals, which represent the single-particle hit position resolution, are shown in Fig.~\ref{fig:residuals} for the innermost tracking station in the north and south arms. 
The position resolution for each of the eight stations varies between 24 and 28~$\mu$m, which is within the design parameters.

\subsection{Electronic Noise}

\begin{figure}[htbp]
	\centering
  \includegraphics[width=8cm]{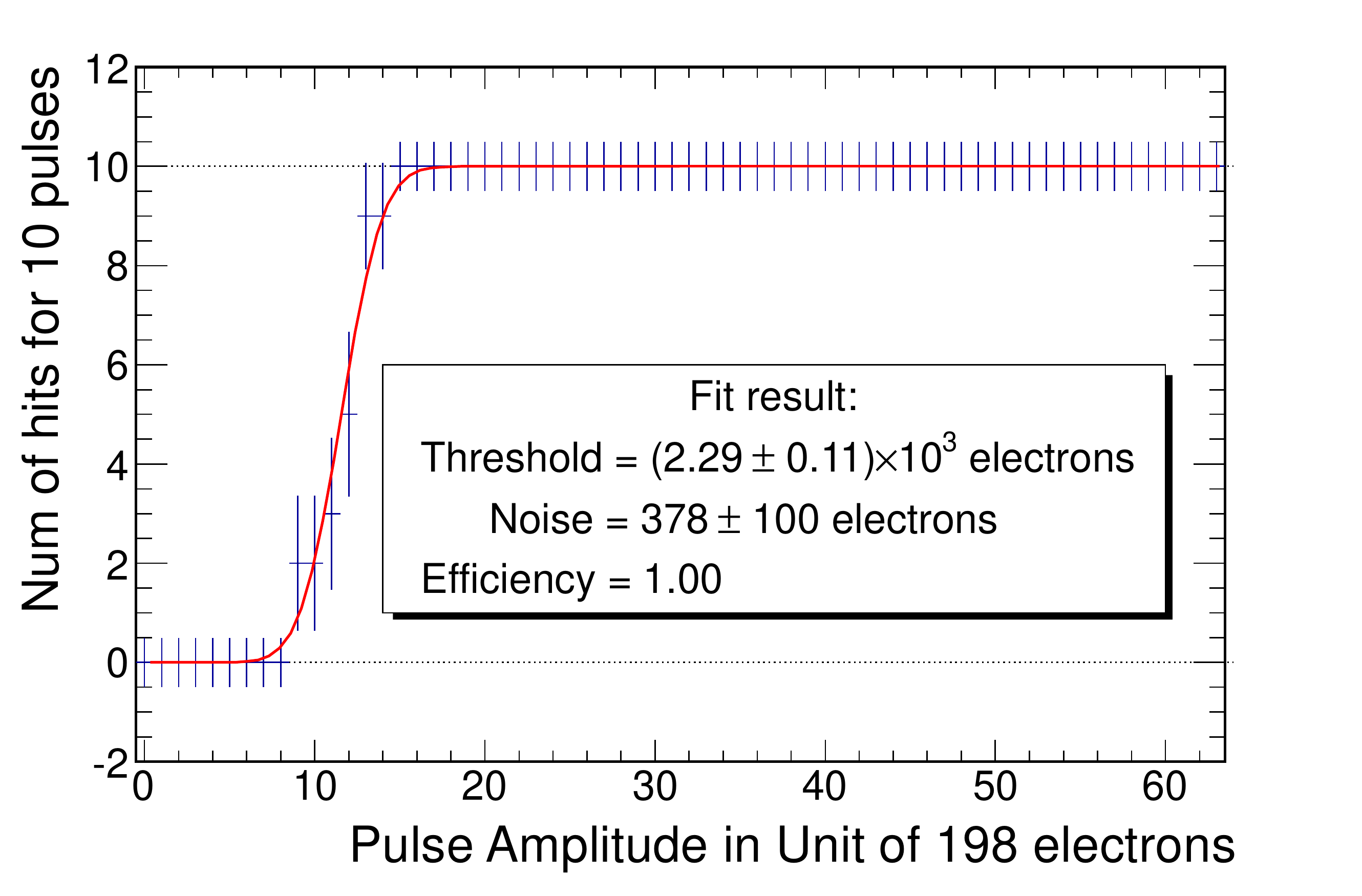}
  \caption{(color online) Typical calibration data for a single channel (blue points), fit with a normal cumulative distribution function.}
	\label{fig:calib}
\end{figure}

The FPHX chip was designed to have a relatively low noise of $\sim$500 electrons when wire bonded
to the actual FVTX sensor (see section \ref{sec:chip}). The electronic noise in the detector is monitored periodically using
the calibration system. During calibration, groups of ten signal pulses of a given height
are sent to an injection capacitor at the front-end of the read-out chip, while the the signal height is scanned across the discriminator
threshold. The noise level is characterized by the broadening of the hit efficiency
threshold as shown in Fig. \ref{fig:calib}. A normal cumulative distribution function is used to fit the data.
The noise level is parametrized by the width, $\sigma$, of the fit function.

\begin{figure}[htbp]
	\centering
  \includegraphics[width=8cm]{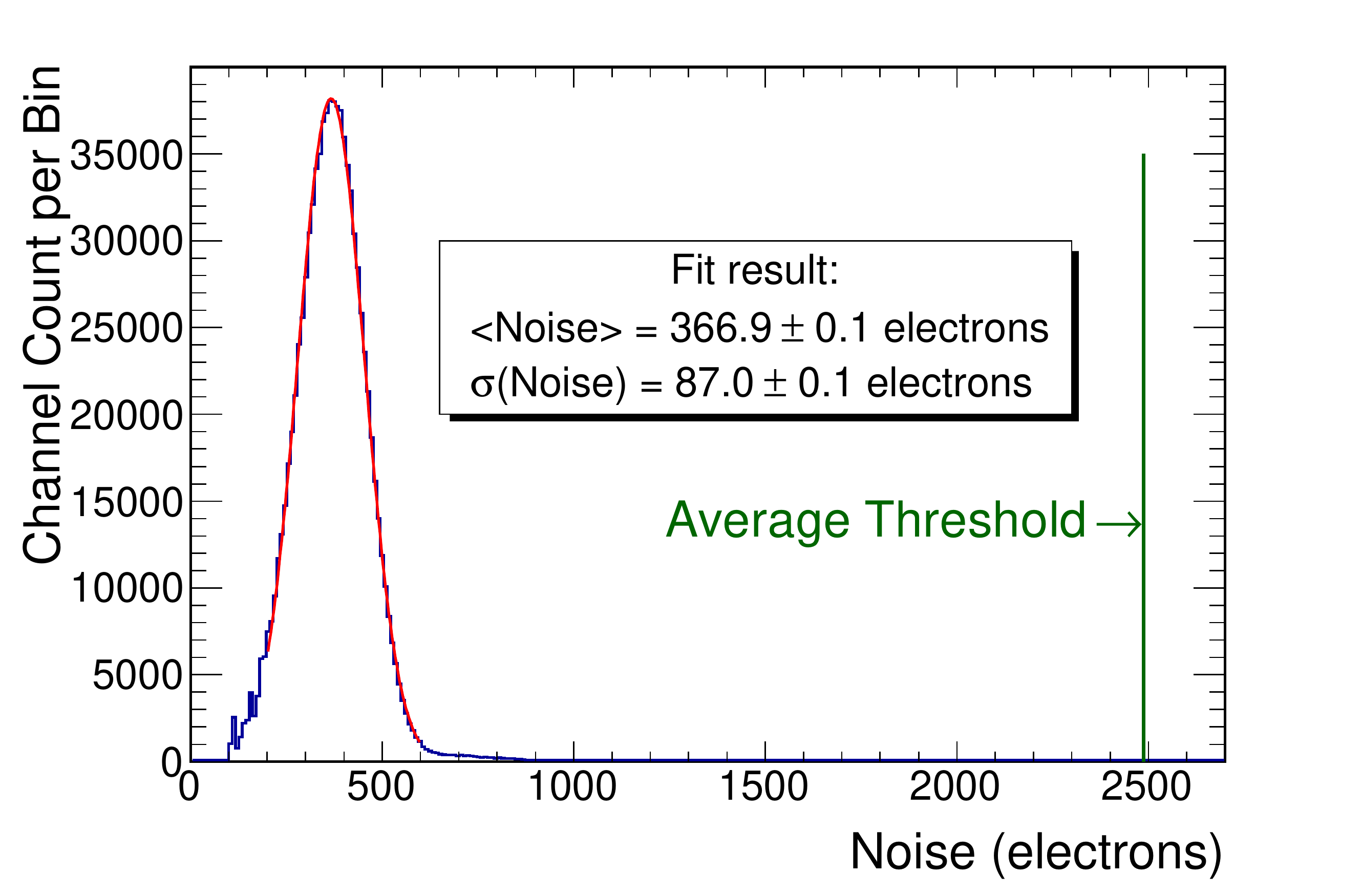}
  \caption{(color online) Histogram of the noise parameter, $\sigma$, for all channels under operating conditions, in a typical calibration run. A Gaussian distribution fit to the data gives a mean noise level of 367 electrons.  The nominal discriminator threshold at \mbox{$\sim2500$} electrons is shown by the vertical green line.}
	\label{fig:all_noise}
\end{figure}

A histogram of the noise level for all operating channels is shown in Fig.~\ref{fig:all_noise}. The
average electronic noise level is between 350-380 electrons, which is significantly lower than the
nominal discriminator threshold of $\sim2500$ electrons.
This level of electronic noise is well within design parameters of the FPHX chip and read-out system.

\section{Summary and Conclusion}
\label{sec:conclusions}

This paper presents a comprehensive report on the design, construction,
and operation of a forward silicon vertex detector, the FVTX, for the
PHENIX experiment at RHIC.  The detector consists of four layers of silicon 
mini-strip sensors at forward and backward rapidity, and enhances 
the capabilities of the existing PHENIX muon arms by providing precision
 tracking of charged particles before they interact in the hadron absorber.  
It was first installed and operated at PHENIX prior to the 2012 RHIC run.  

The detector active area covers the full azimuthal angle over the forward rapidity range $1.2<|\eta|<2.2$.  
Each individual silicon sensor is divided into strips with a 75~$\mu$m pitch in the radial direction 
and a $\phi$ coverage of 3.75$^\circ$.  Groups of custom 128-channel front end read-out ASICs, called FPHX chips, 
are wire bonded to the sensor.  

FVTX sensors display the expected behavior in response to radiation dose.  However, measurements 
of the radiation environment in the PHENIX experimental hall and irradiations of actual FVTX sensors 
show that radiation effects will not significantly degrade detector performance over the expected 
lifetime of the experiment.

From data collected at RHIC, the FVTX has demonstrated single particle hit efficiencies above 95$\%$, single hit position 
resolution better than 30 $\mu$m, and electronic noise levels below 500 electrons, all within design specifications.

The data-push architecture of the FPHX read-out chips allows the FEM to receive all hits registered in the silicon sensors without 
any bias from an external trigger, with minimal processing delay. This feature can potentially allow the FVTX to provide a Level-1
trigger, and can be used to determine the relative luminosity seen by the FVTX detector for each beam crossing.
This is useful in polarized $p+p$ collisions, where the FEM counts all hits, and pairs of coincident hits, above a set ADC threshold
for each beam crossing to determine the relative integrated luminosity for each spin orientation.
Examinations of these new capabilities and analysis of data taken with the FVTX in 2012 and 2013 are underway.%
\section*{Acknowledgments}
\label{sec:ack}  

We gratefully acknowledge the support of the U.S. Department of Energy through the LANL/LDRD Program for this work.  We thank L. J. Bitteker and S. Wender of LANL for assistance with the FVTX 
sensor irradiations at LANSCE. 

\bibliography{main_fvtx}{}
\bibliographystyle{elsarticle-num}
\end{document}